\documentclass[]{raa}            
\usepackage{graphicx,times}
\usepackage{natbib}
\bibpunct{(}{)}{;}{a}{}{,}

\usepackage[a4paper=true,dvipdfm=true,pagebackref=true]{hyperref}
\hypersetup{pdftitle = The title of my PDF, pdfauthor = My name, pdfsubject= The subject, pdfkeywords = keyword1 keyword2 keyword3} 
\hypersetup{colorlinks = true, linkcolor = green, anchorcolor = red, citecolor = blue, filecolor = red, pagecolor = red, urlcolor = red}

\begin{document}

\title{The astrophysical gravitational wave stochastic background}

 \volnopage{ {\bf 20XX} Vol.\ {\bf X} No. {\bf XX}, 000--000}
   \setcounter{page}{1}ep

   \author{T. Regimbau
   }

   \institute{UMR ARTEMIS, CNRS, University of Nice Sophia-Antipolis, Observatoire de la C\^{o}te d'Azur, BP 4229, 06304, Nice Cedex 4, France; {\it regimbau@oca.eu}\\
\vs \no
   {\small Received 2010 Febrary 14; accepted 2010 September 7}
}

\abstract{ A gravitational wave stochastic background of astrophysical origin may have resulted from the superposition of a large number of unresolved sources since the
beginning of stellar activity. Its detection would put very strong
constrains on the physical properties of compact objects, the initial mass
function or the star formation history. On the other hand, it could
be a 'noise' that would mask the stochastic background of
cosmological origin.
We review the main astrophysical processes able
to produce a stochastic background and discuss how it
may differ from the primordial contribution by its statistical
properties. Current detection methods are also presented.
}

   \authorrunning{T. Regimbau}            
   \titlerunning{The Astrophysical Gravitational Wave Stochastic Background }  
   \maketitle

%
%
\section{Introduction}           
\label{sect:intro}

Gravitational wave (GW) astronomy will enable a new window to the Universe
to be opened: not only one expects to discover a set of new exotic
sources, but also to travel back in time, toward the very early
stages of the evolution of the Universe.
According to various cosmological scenarios, we are
bathed in a stochastic background of gravitational waves, memory of
the first instant of the Universe, up to the limits of the Plank
era and the Big Bang \citep{gri01}.
Proposed theoretical models include the amplification of vacuum
fluctuations during inflation \citep{gri74,gri93,sta79}, pre Big Bang models
\citep{gas93,gas03,buo97}, cosmic strings
\citep{vil00} or phase transitions  \cite{cap10} (see \citealt{mag00} or for a general
review and references therein).
In addition to this cosmological background (CGB), an astrophysical
contribution (AGB) may have resulted from the superposition of a large
number of unresolved sources since the beginning of the stellar
activity, which can be either short live burst sources, such as core
collapses to neutron stars \citep{bla96,cow01,cow02a,how04,buo04,mar09,zhu10} or black holes  \citep{fer99a,dara00,dara02a,dara02b,dara04}, oscillation
modes \citep{owe98,fer99b,mar09,zhu10}, final
stage of compact binary mergers \citep{reg06b,reg07}, or periodic long live sources,
typically pulsars \citep{reg01a,reg06a}, the early inspiral phase of compact
binaries \citep{ign01,sch01,far02,coo04} or captures by supermassive black holes \citep{bar04,schn06},
whose frequency is expected to evolve very slowly compared to the
observation time. The nature of the AGB may differ from its
cosmological counterpart, expected to be stationary, unpolarized,
gaussian and isotropic, by analogy with the cosmic microwave
background (CMB).
On the one hand the distribution of galaxies up to 100\,Mpc is not
isotropic but strongly concentrated in the direction of the VIRGO
cluster and the Great attractor; on the other hand, depending whether
the time interval between events is short compared to the duration of
a single event, the integrated signal may result in a continuous, a
popcorn noise or a shot noise background \citep{cow06}.
The optimal strategy to search for a Gaussian (or continuous) stochastic
background is to cross correlate measurements of multiple
detectors, which can be either resonant antennas
such as the cryogenic bars AURIGA, NAUTILUS, EXPLORER, ALLEGRO or
NIOBE \citep{ast90,cer97,pal97,mau96,blai95}, laser interferometers such as LIGO, VIRGO, GEO600,
TAMA300/LCGT and the third generation Einstein Telescope \citep{abr92,bra90,hou92,kur06} on earth or LISA in space \citep{ben98}, or natural detectors such as millisecond
pulsars of the Parkes Pulsar Timing Array (PPTA) \citep{jen05,man06}.
Space and terrestrial detectors will be complementary in the
$10^{-5}-10^4$\,Hz band, while the PPTA is expected to detect GWs at nHz frequencies.
Over the last decade, the first generation of terrestrial detectors
have been built, commissioned and are running in scientific mode at
(or close to) their design sensitivities, providing the opportunity to
do joint data analysis.

This chapter gives an overview of the main features of
the AGBs, from modeling to detection, and discuss how their statistical
properties may differ from those of the cosmological background.
In Section~1, we review the spectral and statistical properties of
astrophysical backgrounds, in Section~2 we introduce the actual detection
method to search for a stochastic background in a network of
detectors, in Section~3 we review the most popular predictions of the
cosmological background, in Section~4 we describe models of AGB and in
Section~5 we discuss the current observational results and future prospects.

\section{The characteristics of the spectrum}

The spectrum of the gravitational stochastic background is usually
characterized by the dimensionless parameter \citep{all99}:
\begin{equation}
\Omega_{\rm gw}(\nu_{\rm o})=\frac{1}{\rho_c}\frac{d\rho_{\rm gw}}{d\ln \nu_{\rm o}},
\label{eq1}
\end{equation}
where $\rho_{\rm gw}$ is the gravitational energy density, $\nu_{\rm o}$ the
frequency in the observer frame and $\rho_c=\frac{3H_0^2}{8 \pi G}$
the critical energy density needed to make the Universe flat
today.
Experimentalists may prefer to work with the spectral
energy density,
\begin{equation}
S_h(\nu_{\rm o})=\frac{3H_0^2}{4 \pi^2} \frac{1}{\nu_{\rm o}^3} \Omega_{\rm gw}(\nu_{\rm o}),
\label{eq2}
\end{equation}
which is directly comparable to the detector sensitivity.

For a stochastic background of astrophysical origin, the energy density
parameter is given by \citep{fer99a}:
\begin{equation}
\Omega_{\rm gw}=\frac{1}{\rho_c c^3} \nu_{\rm o} F_{\nu_{\rm o}},
\label{eq3}
\end{equation}
where the integrate flux received on Earth (in erg/Hz$^{-1}/$/cm$^{-2}$), at the observed frequency $\nu_{\rm o}$ is defined as:
\begin{equation}
F_{\nu_{\rm o}}=\int p(\theta) f_{\nu_{\rm o}}(\theta,z,\nu_{\rm o}) \frac{dR^o(\theta,z)}{dz} d\theta dz, 
\label{eq-flux}
\end{equation}
where $p(\theta)$ is the probability distribution of the source parameters $\theta$.
The first factor in the integral is the fluence of a source (in erg/Hz$^{-1}$/cm$^{-2}$) located at redshift  $z$:
\begin{equation}
f_{\nu_{\rm o}}(\theta,z,\nu_0)=\frac{1}{4 \pi r(z)^2} \frac{dE_{\rm gw}}{d \nu}(\theta,\nu_0(1+z)),
\label{eq5}
\end{equation}
where $r(z)$ is the proper distance, which depends on the adopted cosmology, $\frac{dE_{\rm gw}}{d\nu}(\theta, \nu)$ the gravitational spectral energy emitted and $\nu_{\rm o}(1+z) $ the frequency in the source frame.
The second factor is the number of sources in the interval $\theta-\theta+d\theta$, per unit of time in the observer frame and per redshift interval, is given by:
\begin{equation}
\frac{dR^o(\theta,z)}{dz}=\dot{\rho}^o(\theta,z) \frac{dV}{dz}(z),
\label{eq6}
\end{equation}
where $\dot{\rho}^o(\theta,z)$ the event rate in Mpc$^{-3}$\,yr$^{-1}$ and $\frac{dV}{dz}(z)$ the comoving volume element.

Combining the expressions above, one obtains for the density parameter:
\begin{equation}
\Omega_{\rm gw}(\nu_0)= \frac{8 \pi G}{3 c^2 H_0^3} \nu_{\rm o}
\int d\theta p(\theta) \int^{z_{\sup}}_{z_{\inf}} dz \frac{\dot{\rho}^o(\theta,z)}{E(\Omega,z)}
\frac{dE_{\rm gw}}{d\nu}(\theta,\nu_0(1+z)) .
\label{eq7}
\end{equation}
Replacing the constants by their usual values we get:
\begin{equation}
\Omega_{\rm gw}(\nu_{\rm o})= 5.7 \times 10^{-56}(\frac{0.7}{h_0})^2  \nu_{\rm o}
\int d\theta p(\theta) \int^{z_{\sup}}_{z_{\inf}} dz \frac{\dot{\rho}^o(z)}{E(\Omega,z)} \frac{dE_{\rm gw}}{d\nu}(\theta,\nu_0(1+z)), 
\label{eq-omega}
\end{equation}
where $\dot{\rho}^o$ is given for $h_0=0.7$.
The limits of the integral over $z$ depend on both the emission frequency range in the source frame, and the redshift interval, where the source can be located:
\begin{equation}
z_{\sup} (\theta,\nu_{\rm o})=
\left\lbrace
\begin{array}{ll}
z_{\max}    &   \hbox{  if } \nu_{\rm o} < \frac{\nu_{\max} }{(1+z_{\max})}\\
\frac{\nu_{\max}}{\nu_{\rm o}}-1 &   \hbox{  otherwise }\\
\end{array}
\right.
\label{eq-zsup}
\end{equation}
and \begin{equation}
z_{\min} (\theta,\nu_{\rm o})=
\left\lbrace
\begin{array}{ll}
z_{\min}    &   \hbox{  if } \nu_{\rm o} > \frac{\nu_{\min} }{(1+z_{\min})}\\
\frac{\nu_{\min}}{\nu_{\rm o}}-1 &   \hbox{  otherwise }\\
\end{array}
\right.
\label{eq-zinf}
\end{equation}
Consequently, the shape of the spectrum of any astrophysical background is
characterized by a cutoff at the maximal emission frequency and a
maximum at a frequency which depends on the shape of both the redshift distribution and
the spectral energy density.

For most of the models presented in Section~5, the event rate per unit of redshift can be derived directly from the cosmic star formation rate. In the simple case when the gravitational emission occurs shortly after the birth of the progenitor, it is given by:
\begin{equation}
\dot{\rho}^o(\theta,z)= \lambda(\theta,z) \frac{\dot{\rho}_*(z)}{1+z}, 
\label{eq-rate}
\end{equation}
where $\lambda$ is the mass fraction converted into the progenitors in M$_{\odot}^{-1}$, which depends on the initial mass function,  $\frac{dV}{dz}$ the element of comoving volume and $\dot{\rho}_*(z)$  the cosmic star formation rate (SFR) in M$_{\odot}$\,Mpc$^{-3}$\,yr$^{-1}$. The  $(1+z)$ factor in the denominator corrects for the time dilatation due to the cosmic
expansion.

\begin{figure}[ht!]
\centering
\includegraphics[angle=0,width=0.8\columnwidth]{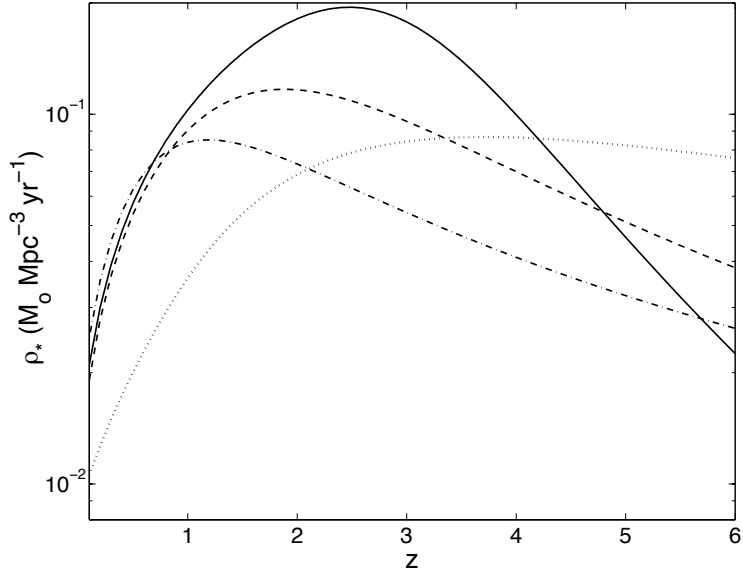}
\caption{Cosmic star formation rates (in M$_\odot$\,Mpc$^{-3}$\,yr$^{-1}$) used in this paper: \cite{hop06} 
(continuous line), \cite{far07} (dashed line),
\cite{wil08} (dot-dashed line), and the fossil
model of \cite{nag06} (dot line).  As discussed in the text,
these rates are largely the same up to $z \sim 1$, but show important
differences at higher redshift.
\label{fig-sfr}}
\end{figure}

Observations of star forming galaxies
with large telescopes such as the Keck or the Hubble Space Telescope
have extended our view of the Universe up to redshifts $z \sim 5-6$, by
tracing the evolution with cosmic time of the galaxy luminosity
density.
The main uncertainty comes from dust extinction, which
spreads the UV luminosity into the far IR. \cite{mad98} derived an
expression that matches most of the measurements in the U-V continuum and
H$\alpha$, up to $z \sim 4$, and that includes an extinction correction of
A$_{1500}$=1.2\,mag. The SFR is expected to increase rapidly
between $z \sim 0-1$, peak around $z \sim 1.7$ and smoothly decrease
at large redshifts. After $z \sim 2$, the behavior must be regarded
as tentative, due to the large uncertainties in the estimates of the
U-V luminosity from the Lyman break galaxies in the Hubble Deep Field.
\cite{ste99} proposed another scenario where the SFR remains constant after
$z \sim 2$. Other studies suggested even an increase of the SFR,
claiming that it may have been severely underestimated due to large amount of dust extinction \citep{bla99}.
However, the hypothesis of a gentle decline at high redshifts seems to
be favored by new measurements of the galaxy luminosity function in the
UV (SDSS, GALEX, COMBO17) and FIR wavelengths (Spitzer Space
Telescope), which allowed to refine the previous models of star
formation history, up to redshift $z \sim 6$, with tight constraints
at redshifts $z<1$.
In a recent work, \cite{hop06} used the Super Kamiokande limit on the
electron antineutrino flux from past core-collapse supernovas
to derive parametric fits of the form of \cite{col03}.
Investigating the effect of the initial mass
function (IMF) on the normalization of the SFR, they showed
that top heavy IMFs are preferred to the traditional Salpeter IMF
\citep{sal95}, and the fits are optimized for IMFs of the form:
\begin{equation}
\xi(m) \: \propto
\left\lbrace
\begin{array}{ll}
(\frac{m}{m_0})^{-1.5}    &   \hbox{  for } 0.1<m<m_0 \\
(\frac{m}{m_0})^{-\gamma} &   \hbox{  for } m_0<m<100 \\
\end{array}
\right.
\label{eq-imf}
\end{equation}
with a turnover below $m_0=0.5$ M$_{\odot}$, normalized within the
mass interval $0.1 - 100$ M$_{\odot}$ such as $\int m\xi(m)dm$ = 1, and with  $\gamma = 2.35$ (A modified Salpeter).
\cite{far07} used a different set of measurements and different dust extinction
corrections and found an SFR similar to  that of
\cite{hop06} up to $z \sim 1$, but which decreases slightly at higher
redshifts. \cite{wil08} used measurements of the stellar
mass density and derived an SFR  equivalent to that of \cite{hop06,far07} for
redshifts smaller than $z \sim 0.7$, but again is lower at higher redshifts.  
Finally, \cite{nag06}, derived a model from the
fossil record of star formation in nearby galaxies.  It is probably
underestimated at small redshifts, and is constant at high redshifts
due to the contribution of elliptical galaxies.  Note that at present
there is a discrepancy between the ``instantaneous'' SFR, measured
from the emission of young stars in star forming regions, and the SFR
as determined from extragalactic background light.  This could have an
important impact on the contribution to the confusion background for
sources from $z > 2$. However, it shouldn't noticeably affect the results, since sources beyond
$z \sim 2$ are too weak to contribute significantly to the integrated
signal.
Figure~\ref{fig-sfr} compares the four prior models described above, calculated for the
flat Einstein de Sitter 737 cosmology, with $\Omega_m=0.3$,
$\Omega_{\Lambda}=0.7$ and Hubble parameter $H_0=70$ km s$^{-1}$
Mpc$^{-1}$ \citep{rao06}, corresponding to the so-called concordant
model derived from observations of distant type Ia supernovae
\citep{per99}  and the power spectra of the cosmic microwave background
fluctuations \citep{spe03}.

Besides the spectral properties, it is important to study the nature of
the background. 
In the case of short-lived signals, they may show very different statistical behavior depending on the ratio between the duration of the events and the time interval between successive events, the duty cycle:
\begin{equation}
\Delta(z)=\int^z_0 \bar{\tau} (1+z') \frac{dR^o(z')}{dz'} dz'
\label{eq-DC}
\end{equation}
which is also the average number of events present at the detector at a given observation time.

\textbf{Continuous:} the number of sources is large enough for the time interval 
between events to be small compared to the duration of a single event.
The waveforms overlap to create a continuous background and 
due to the central limit theorem, such backgrounds obey the Gaussian 
statistic. They are completely determined by their spectral properties and 
could be detected by data analysis methods in the frequency domain such as 
the standard cross correlation statistic \citep{all99}. 

\textbf{Shot noise:} the number of sources is small enough for the time interval 
between events to be long compared to the duration of a single event.  The 
waveforms are separated by long stretches of silence and the 
closest sources may be detected by data analysis techniques in the time 
domain (or the time frequency domain) such as match filtering \citep{arn99,pra01}. 

\textbf{Popcorn:} an interesting intermediate case arises when the time 
interval between events is of the same order of the duration of a single 
event. These signals, which sound like crackling popcorn, are known as 
"popcorn noise". The waveforms may overlap but the statistic is not 
Gaussian anymore and the amplitude on the detector at a given time is 
unpredictable.   
Promising data analysis strategies have been investigated in the last few years, such as the Maximum Likelihood statistic, an extension of the cross correlation statistic in the time domain
\citep{dra03} or methods based on the Probability Event Horizon concept
\citep{cow05}, which describes the evolution of the cumulated signal throughout the Universe, as a function of the
observation time.

\begin{figure}
\centering
\includegraphics[angle=0,width=1\columnwidth]{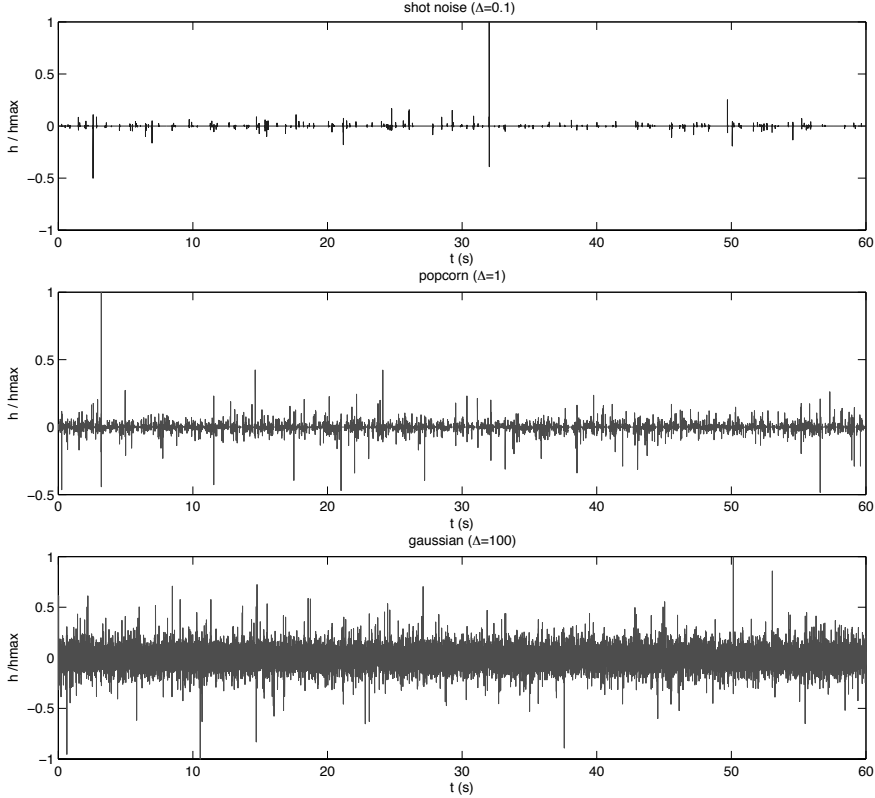}
\caption{time series corresponding to shot noise, popcorn and gaussian regimes}
\label{fig-series}
\end{figure}
\section{Detection}

The optimal strategy to search for a gaussian (or continuous) stochastic
background, which can be confounded with the intrinsic noise
background of the instrument, is to cross correlate measurements of multiple
detectors.
In this section, we give a brief overview of the standard data analysis
technique currently used for terrestrial interferometers.

\begin{figure}[ht!]
\centering
\includegraphics[angle=0,width=0.7\columnwidth]{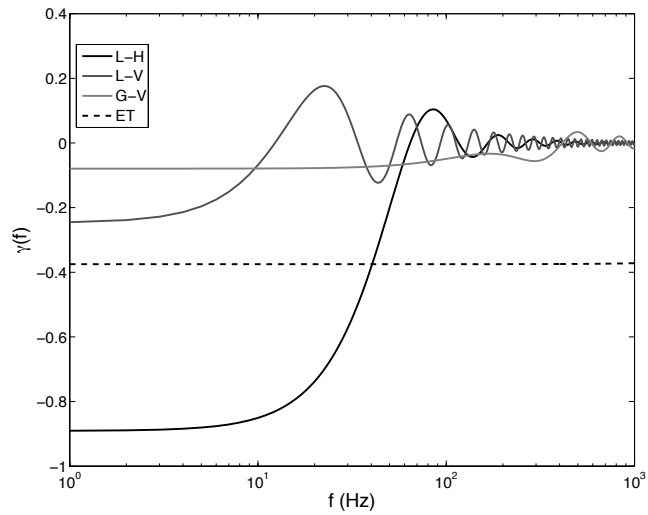}
\caption{overlap reduction function for the most promising detector pairs. L stands for LIGO Livingston and H for LIGO Hanford, V for Virgo, G for GEO600 and ET for the planned Einstein Telescope in the triangular configuration.
\label{fig-overlap}}
\end{figure}

When the background is assumed to be isotropic,
unpolarized and stationary, the cross correlation product is
given by \cite{all99}:
\begin{equation}
Y=\int_{-\infty}^\infty \tilde{s_1}^*(f)\tilde{Q}(f)\tilde{s_2}(f) df,
\end{equation}
where
\begin{equation}
\tilde{Q}(f)\propto \frac{\Gamma (f) \Omega_{\rm gw}(f)}{f^3P_1(f)P_2(f)}
\end{equation}
is a filter that maximizes the signal to noise ratio ($S/R$). In the
above equation, $P_1(f)$ and $P_2(f)$ are the power spectral noise
densities of the two detectors and $\Gamma$ is the non-normalized overlap
reduction function, characterizing the loss of sensitivity due to
the separation and the relative orientation of the detectors (Fig.~\ref{fig-overlap},
The optimized $S/N$ ratio for an integration time $T$ is given by \cite{all97}:
\begin{equation}
(\frac{S}{N})^2 =\frac{9 H_0^4}{8 \pi^4}T\int_0^\infty
df\frac{\Gamma^2(f)\Omega_{\rm gw}^2(f)}{f^6 P_1(f)P_2(f)}.
\end{equation}

\begin{figure}[hb!]
\centering
\includegraphics[angle=0,width=0.7\columnwidth]{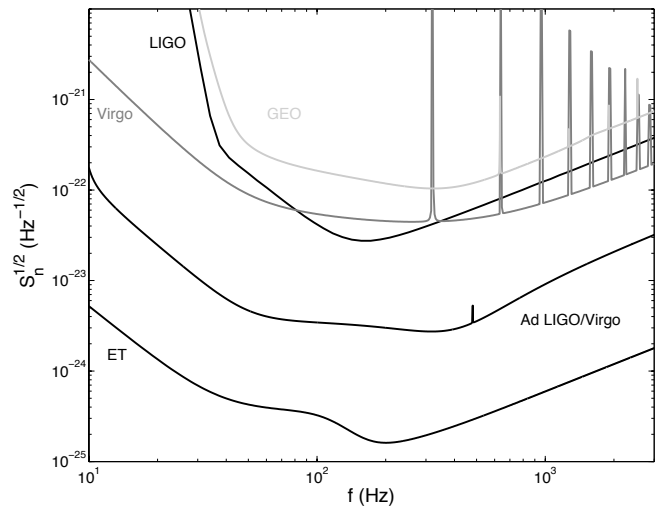}
\caption{designed sensitivities of the main first generation
  interferometers, compared to the planned sensitivities
  of advanced detectors (LIGO or Virgo) and third generation detectors (Einstein Telescope).
\label{fig-noise}}
\end{figure}

In the literature, the sensitivity of a pair of detectors is usually given
in terms of the minimum detectable amplitude corresponding to
$\Omega_{\rm gw}$ equal to constant (hereafter flat spectrum) \citep{all99}:
\begin{equation}
\Omega_{\min}=\frac{4 \pi^2}{3H_0^2\sqrt{T}}({\rm erfc}^{-1}(2 \beta)-{\rm erfc}^{-1}(2 \alpha)) \lbrack \int_0^\infty df \frac{\Gamma^2 (f)}{f^6 P_1(f)
 P_2(f)}\rbrack^{-1/2}.
\end{equation}
The expected minimum detectable amplitudes for the main
terrestrial interferometer pairs, at design sensitivity (Fig.~\ref{fig-noise}, and after one
year of integration, are given in
Table~\ref{table-sensitivity}, for a detection rate $\alpha=90\%$ and
a false alarm rate $\beta=10\%$. 

\begin{table}
\caption{Expected $\Omega_{\min}$ for the main detector pairs,
  corresponding to a flat background spectrum, one year of integration over the full frequency band, a detection rate $\alpha=90 \%$  and a false alarm rate
  $\beta=10\%$. LHO and LLO stand for LIGO Hanford Observatory and
  LIGO Livingston Observatory, ET stand for ET for the planned Einstein Telescope in the triangular configuration \citep{fla93}.}
\label{table-sensitivity}
\centering
\begin{tabular}{lcccc}
\noalign{\smallskip}
\hline
\noalign{\smallskip}
& LHO-LHO & LHO-LLO & LLO-VIRGO & VIRGO-GEO  \\
\noalign{\smallskip}
\hline
\noalign{\smallskip}
initial& $4 \times 10^{-7}$ & $3 \times 10^{-6}$ &  $6 \times 10^{-6}$
&  $2 \times 10^{-5}$  \\
advanced & $6 \times 10^{-9}$  & $1 \times 10^{-9}$  & &  \\
\noalign{\smallskip}
\hline
\noalign{\smallskip}
ET& $5 \times 10^{-12}$ &&& \\
\noalign{\smallskip}
\hline
\end{tabular}
\end{table}

$\Omega_{\min}$ is of the order of $10^{-6}-10^{-5}$ for the first
generation of interferometers combined as LIGO/LIGO and
LIGO/Virgo. Their advanced counterparts will permit an increase of two
or even three orders of magnitude in sensitivity ($\Omega_{\min} \sim
10^{-9}-10^{-8}$). The pair formed by the co-located and co-aligned
LIGO Hanford detectors, for which the overlap reduction function is
equal to one, is potentially one order of magnitude more sensitive
than the Hanford/Livingston pair, provided that instrumental and environmental
noises can be removed.
In Table~\ref{table-upperlimits} we show the evolution of the upper limit obtained with the LIGO detectors in a narrow band around 100\,Hz, and corresponding to $\Omega_{\rm gw}$ equal to constant at all frequencies. In Table~\ref{table-upperlimits2}, the latest published LIGO upper limit is compared to observational limits already achieved with resonant  bar experiments at about 900\,Hz and pulsar timing at nHz frequencies .

\begin{table}
\caption{evolution of the LIGO 90\% Bayesian upper limit on a frequency independent
  $\Omega_{\rm gw}$.} 
\label{table-upperlimits}
\centering
\begin{tabular}{lcccc}
\noalign{\smallskip}
\hline
\noalign{\smallskip}
Run & frequency band (Hz) & upper limit & reference\\
\noalign{\smallskip}
\hline
\noalign{\smallskip}
S1 &  $40-314$  & 23 &\cite{S1}\\
S3 &  $69-156$  & $8.4 \times 10^{-4}$ &\cite{S3}\\
S4 &  $51-150$ & $6.5 \times 10^{-5}$ &\cite{S4}\\
S5 &  $40-170$  & $5.9 \times 10^{-6}$ &\cite{S5}\\
\noalign{\smallskip}
\hline
\end{tabular}
\end{table}

\begin{table}
\caption{best published direct upper limits on a frequency independent
  $\Omega_{\rm gw}$ derived from correlation, for different type
  of experiments.}
\label{table-upperlimits2}
\centering
\begin{tabular}{lcccc}
\noalign{\smallskip}
\hline
\noalign{\smallskip}
Type of detectors & Experiment & frequency (Hz) & upper limit & reference\\
\noalign{\smallskip}
\hline
\noalign{\smallskip}
Room Temp. resonant bars & Glasgow &  985 & 6125 &\cite{hou75}\\
Cryogenic resonant bar & Explorer+Nautilus & 907 & 120 &\cite{ast99} \\
Pulsar timing & Parkes & $~10^{-8}$ & $4 \times 10^{-8}$  &\cite{jen05}\\
\noalign{\smallskip}
\hline
\end{tabular}
\end{table}

An extension of the cross-correlation method to non isotropic contributions has been investigated by \cite{all97b,cor01,bal06} and \cite{mit08}. The basic idea is to use multiple detector pairs to create maps of anisotropy of the GW background, similar to a radiometer of GWs.

\section{Relic stochastic background}

Mechanisms able to generate stochastic backgrounds of GWs in the very
early stages of the Universe have been investigating intensively in the
past decades. Their detection would have a profound impact on our
understanding of near Big Bang cosmology and high energy physic,
providing a unique way to explore the Universe a fraction of second
after the Big Bang, after gravitons decouple from the primordial plasma.
It is not the purpose of this article to develop in details all the different models of cosmological stochastic background present in the literature, as our main interest is the astrophysical background, but for comparison purpose, we give in this section a rapid overview of some popular predictions that could be masked by the astrophysical background. We refer interested readers to very nice review papers by
\cite{all97,mag00} and \cite{buo03} .

In this section, unless it is mentioned otherwise, the
Hubble parameter is assumed to be $H_0=70$\,km\,s$^{-1}$\,Mpc${^-1}$.
\begin{figure}
\centering
\includegraphics[angle=0,width=0.8\columnwidth]{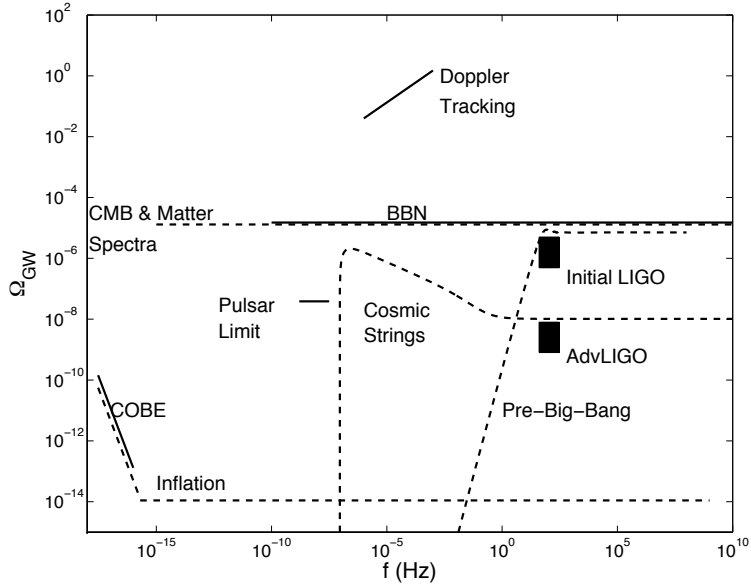}
\caption{Theoretical predictions of the cosmological stochastic
  background and observational bounds. The cosmic strings plot
  corresponds to $p=0.1$, $\varepsilon=7 \times 10^{-5}$, $G
  \mu=10^{-8}$. This figure was kindly provided by Vuk Mandic.
\label{fig-cgb}}
\end{figure}

\subsection{Amplification of Vacuum Fluctuations}
Amplifications of vacuum fluctuations at the transitions between the
de Sitter, radiation dominated (RD) and matter dominated (MD) eras,
first discussed by \cite{gri74,gri93} and \cite{sta79}, are expected to produce a
GW background which spectrum and amplitude depend strongly on the
fluctuation power spectrum developed during the early inflationary
period. The standard de Sitter inflation predicts a spectrum that
decreases as ~1/$f^2$ in the range $3 \times 10^{-18}-10^{-16}$\,Hz and
then remains constant in a very large band up to MHz frequencies. In the
low frequency region, modes amplified during inflation at both the de
Sitter-RD and RD-MD transitions contribute. The turnover between the
two phases corresponds to the limit at which only the modes amplified during
the de Sitter-RD transition can be observed.
The COBE experiment, which has the same $1/f^2$ behaviour in the low
frequency region, provides an upper bound
of $\Omega_{\rm gw} \sim 9 \times 10^{-14}$ \citep{mag00,buo03} for the
flat region. Actually, a GW background larger than $\Omega_{\rm gw} \sim 7
\times 10^{-10} (\frac{10^{-18}}{f})^2$ at frequencies between $3 \times
10^{-18}-10^{-16}$\,Hz would have produced stochastic frequency
redshifts through Sachs-Wolfe effects, which would have resulted
into temperature fluctuations larger than those measured for the
cosmic microwave background.
In the more realistic
scenario of ``slow roll down'' inflation, the inflaton field rolls
toward the minimum of its potential, producing an acceleration of the
expansion. The Hubble parameter
is not constant like in the standard scenario, but
decreases monotonically during the period of inflation. GWs
are produced by fluctuations that go out the Hubble radius during
inflation, and re-enter at the radiation era. The resulting spectrum
is not flat as for the de Sitter inflation but rather has a $f^{n_T}$
dependency, where $n_T<0$ and $|n_T|<<1$.
The spectral index can be expressed in terms of the
scalar and tensorial contributions to the quadrupole cosmic microwave
background (CMB) anisotropy as $n_T=-T/7S$ \citep{mag00}.
The most optimistic predictions for a detection with LISA at $f \sim 10^{-4}$, corresponding to $n_T=0.175$, give an amplitude $\Omega_{\rm gw} \sim 10^{-15}$, but $n_T$ could be much smaller, of the order of $10^{-3}$ \citep{mag00}. 
In a recent paper, \cite{ton09} studied the effect of a running spectral index $\alpha_t$ on the GW spectrum and found that $\alpha_t >0$ could enhance the signal significantly, especially at high frequencies.

A more interesting case arises from pre-big-bang scenarios in string
cosmology \citep{gas93,gas03}. According to these models, the standard RD and MD eras were
preceded by phases in which the Universe was first large and
shrinking (inflaton phase) and then characterized by a high curvature
(stingy phase). The GW spectrum produced at the transition between the
stingy phase and the RD era is described as $\Omega_{\rm gw}\sim f^3$ for
$f<f_s$ and $\Omega_{\rm gw} \sim f^{3-2\mu}$ for $f_s<f<f_1$ \citep{buo97,mand06}.
The turnover frequency is essentially unconstrained, $\mu<1.5$
reflects the evolution of the Universe during the 'stingy' phase and
the cutoff frequency $f_1$, which depends on string related
parameters, has a typical value of $4.3 \times 10^{10}$\,Hz. An upper
limit on $\Omega_{\rm gw}$ is imposed by the Big Bang
Nucleosynthesis (BBN) bound down to $10^{-10}$\,Hz, corresponding to the
horizon size at the time of BBN. Actually, if the total energy amount
carried by GWs, $\int \Omega_{\rm gw} d(ln f)$, at the time of
nucleosynthesis was larger than $1.1 \times 10^{-5} (N_{\nu} -3)$,
where $N_{\nu}$ is the effective relative number of species at the
time of BBN, it would have resulted into a particle production rate too
large compared to the expansion of the Universe, to account for the primordial
abundances of the light elements $^2$H, $^2$He, $^4$He and $^7$Li.
Measurements of the light element abundances combined with the WMAP
data gives $N_{\nu}<4.4$ \citep{cyb05}, which translates to $\Omega_{\rm gw} < 1.5
\times 10^{-5}$. Recent measurements of CMB anisotropy spectrum, galaxy power spectrum
and of the Lyman-$\alpha$ forest give a bound of similar amplitude
which extends down to $10^{-15}$\,Hz, corresponding to the
horizon size at the time of CMB decoupling \citep{smi06}.

\subsection{Cosmic Strings}
Cosmic strings, formed as linear topological defects during
symmetry breaking phase transitions or in string theory inspired
inflation scenarios, may emit GWs by oscillating
relativistically and shrinking in size \citep{buo03}.
CMB observations are not consistent with the most promising scenario of
very large mass-per-unit-length strings, acting as initial seeds for the
formation of large-scale structures at the GUT scale symmetry
break, but strings of lower energy scale may still contribute to
the CGB. Also, in models with non vanishing cosmological constant, it
can still be a viable option \citep{ave98,bat98}.
The spectrum is expected to peak around the frequency $f \sim
10^{-12}$ and become almost flat in a large frequency band from
$10^{-8}$ to $10^{10}$\,Hz, where the amplitude can reach
$\Omega_{\rm gw} \sim 10^{-9}-10^{-8}$, according to numerical simulations
of a cosmic strings network with $G \mu<10^{-6}$ \citep{buo03}.
At the present time, the most stringent constrain is given by pulsar timing observations.
When passing between Earth and pulsars, GWs may cause fluctuations
in the time arrival of the pulses. Observations of PSR B1805+09
\citep{kas94,lom03}, a very stable narrow profile millisecond pulsar, with
the Arecibo and Green Bank radiotelescopes for over 17 years, give an upper limit
of $\Omega_{\rm gw} \sim 1.2 \times 10^{-9}$  at the frequency
$f=\frac{1}{T_{\rm obs}}=1.86 \times 10^{-9}$\,nHz, and combining the timings of
seven pulsars citep{jen05} placed a lower bound of $\Omega_{\rm gw} \sim 4
\times 10^{-8}$.
The Parkes Pulsar Timing Array project \citep{man06} which is expected to
reach $\Omega_{\rm gw} \sim 2 \times 10^{-10}$, by  monitoring
twenty pulsars for five years, may be our best hope to detect cosmic
strings in the near future.

In a recent work, \cite{dam00,dam01,dam05} and \cite{sie06} considered the stochastic
background created by cusps of oscillating cosmic superstring loops at
the end of Brane inflation.
The amplitude and the shape of the GW spectrum is determined by three
parameters \citep{dam05}: the string tension $\mu$, the reconnection
probability $p$, typically in the range $10^{-3}-1$ and $\varepsilon$
the typical size of the closed loops produced in the string network.
In particular, $\mu$ and $\varepsilon$ determine the lowest frequency
at which a string loop can emit GWs.
The GW spectrum is characterized by a decrease at lowest frequencies,
followed by a flat region. Assuming $p=5 \times 10^{-3}$, $G \mu=10^{-7}$
and $\varepsilon=10^{-7}$, \citep{sie06} obtained a spectrum that avoids the low
frequency bound due to CMB or pulsar timing measurements but still
remain in the sensitivity band of space or ground based detectors.
Let's mention that according to \cite{dam00,dam01} and \cite{dam05} occasional
strong beams of GWs could be produced at cusps, forming a popcorn like
noise on top of the gaussian contribution.

\subsection{Phase Transitions}

At the early stages of its evolution, the Universe may have undergone several episodes of phase
transition, in which the symmetry of particle-physics fundamental
interactions spontaneously broke. This may occur for instance at the
QCD (150\,Mev) and electroweak scales (100\,GeV) or even earlier, at the
grand unified scale (see \citealt{mag00} and references therein).
The Standard Model predicts rather a smooth crossover, but in its
supersymmetric extensions, the transition from a metastable phase (the false vacuum) to the state of broken
symmetry (the true vacuum) can be first order, and large amount of GWs could be produced when bubbles
of the new phase are nucleated, grow and as they become more numerous, collide at very high velocities.
The GW spectrum reach a maximum of $\Omega_{\rm gw} \sim 10^{-6}
(\frac{H_*}{\beta})^2(\frac{100}{N_*})^{1/3}$ at $f_{\rm max} \sim 3
\times (\frac{\beta}{H_*})(\frac{N_*}{100})^{1/6}T_*$,
where $\Gamma=\Gamma_0 e^{-\beta t}$ is the nucleation rate of
bubbles, $T_*$ is the temperature in GeV of the phase transition, $H_*$ the relevant Hubble parameter and $N_*$ the number of relativistic degrees of freedom.
In particular, a phase transition at the electroweak scale could give
a detectable signal of $\Omega_{\rm gw} \sim 10^{-9}$ at the mHz frequency, where LISA is the most sensitive.
Besides the collision of the broken phase bubbles, other processes are expected to produce gravitational waves, such as the magnetohydrodynamical turbulence in the plasma stirred by the bubble collisions, and the magnetic fields amplified by the magnetohydrodynamical turbulence \citep{cap10}.

\section{Sources of Astrophysical backgrounds}

Many examples of astrophysical backgrounds can be found in the
literature. However, a
direct comparison between the different models is made difficult by
the fact that they often use different cosmologies, SFRs, IMFs, or
mass range for neutron star (NS) or black hole (BH) progenitors.
In this section we review some of the most promising predictions,
since it is impossible to cover in one chapter all the literature on the
subject.

\subsection{Binary Neutron Stars}

Double neutron star coalescences, which may radiate about $10^{53}$ erg in the last seconds of their inspiral trajectory, up to $1.4-1.6$\,kHz, may be the most important contribution in the frequency range of ground based detectors \citep{reg06b,reg08}.

The coalescence rate per comoving volume $\dot{\rho}^{\rm o}$ in Equation~\ref{eq-omega}, results from the convolution of the formation rate of the progenitors 
with the probability distribution $P$ of the delay $t_d$  between the formation of the progenitors and the coalescence:
\begin{equation}
\dot\rho^o_{c}(z) \propto \int \frac {\dot\rho_*(z_f)} {(1+z_f) } P(t_d)dt_d,
\end{equation}
where $z$ is the redshift at the time of the coalescence and $z_f$  is the redshift at the time of formation of the binary. 
Population synthesis \citep{pir92,tut94,lip95,and04,dfp06,bel06,sha08} suggest that the delay time is well described by a probability distribution of the form:
\begin{equation}
P_d(t_d) \propto \frac{1}{t_d} \,\ \mathrm{with} \,\ t_d > \tau_0.
\end{equation}
This broad model accounts for the wide range of merger times observed in binary pulsars and is also consistent with short gamma ray burst  observations in both late and early type galaxies \citep{ber06}.   
\cite{bel01} and \cite{bel06} have identified a new efficient formation channel which produces a significant fraction of tight binaries with merger times in the range $\tau_m \sim 0.001-0.1$\,Myr, which gives a minimal delay time $\tau_0 \sim 20$\,Myr, corresponding roughly to the time it takes for massive binaries to evolve into two neutron stars. 

\begin{table}
\caption{\label{table-rates} Taken from Table 4 of \cite{pos06}, most current estimates of the Galactic merger rates of NS-NSs and NS-BHs, derived from statistical studies (first row), and from population synthesis. The high rate obtained by Tutunov and Yungelson (1993) is due to the assumption that neutron stars or black holes are born with no kick velocity, leading to an overestimate of  the number of systems that survive the two supernovae. The low rate obtained by Voss and Tauris (2003) is due to the use of a different value of the parameter $\lambda$, which measures the binding energy of the common envelop.}
\centering
\begin{tabular}{lcc}
\hline\hline
\bf{statistics} & NS-NS &\\
Kalogera et al. (2004) & 83 (17-292) &\\
\hline
\bf{population synthesis} & NS-NS &  NS-BH \\
Tutunov and Yungelson (1993) & 300 & 20 \\
Lipunov et al. (1997) & 30 & 2 \\
Potergies Zwart and Yungelson (1998) & 20 & 2 \\
Nelemans et al. (2001) & 20 & 4 \\
Voss and Tauris (2003) & 2 & 0.6 \\
O'Shaughnessy et al. (2005) & 7 & 1 \\
de Freitas Pacheco et al. (2006) & 17  \\
Belczinsky et al. (2007) & 10-15 & 0.1 \\
O'Shaughnessy et al. (2008) & 30 & 3 \\\hline
\end{tabular}
\end{table}
The local cosmological rate at $z=0$, $\dot{\rho}_{\rm o}$ in  Myr$^{-1}$\,Mpc$^{-3}$, is usually extrapolated by taking the product of the rate in the Milky Way ($r_{mw}$ in yr$^{-1}$) and the density of Milky-Way equivalent galaxies, given from measurements of  the blue stellar luminosity around $n_{\mathrm{mw}} \sim (1-2) \times 10^{-2}$\,Mpc$^{-3}$ \citep{phi91,kal01,kop08}.
The most current estimates of the NS-NS galactic coalescence rate are given in the range $1-817$\,Myr$^{-1}$ ($95 \%$ confidence intervals) for  statistical studies which extrapolate the rates from observed galactic NS-NS \citep{kal04}, preferably between $17-292$ ($95 \%$ confidence intervals) with a peak probability around $83$\,Myr$^{-1}$, and in the range $1-300$, more likely around $10-30$,  for population synthesis models, which combine theoretical and observational constraints (Table~\ref{table-rates}). 
In the quadrupolar approximation, the GW energy spectrum emitted by a binary system, which inspirals in a circular orbit is given up to the last stable $\nu_{\max}$ orbit by:
\begin{equation}
dE_{\rm gw}/{d\nu} = \frac{(G \pi)^{2/3}}{3} \frac{m_1m_2}{(m_1+m_2)^{1/3}} \nu^{-1/3}.
\end{equation}
Assuming $m_1=m_2=1.4$ M$_\odot$ for the star masses,  the energy density increases as $\nu_o^{2/3}$ before it reaches a maximum of $\Omega_{gw} \sim 3.5 \times 10^{-9} \dot{\rho}_0$ at around 500 Hz, where $\dot{\rho}_0$ is the local rate in My$^{-1}$ Mpc$^{-3}$ (about 0.01 times the galactic rate).
This means that ET should be able to detect the background from binaries even for the most pessimistic predictions of the coalescence rate, down to $\dot{\rho}_0 \sim 0.02$ (roughly equivalent to a galactic rate of 2 My$^{-1}$), for a signal-to-noise ratio of 3, after one year of observation.

\subsection{Rotating Neutron Stars: Tri-axial Emission}
Rotating neutron stars with a triaxial shape may have a time varying
quadrupole moment and hence radiate GWs at twice the rotational
frequency.
The total spectral gravitational energy emitted by a neutron star born with a rotational period $P_0$, and
which decelerates through magnetic dipole torques and GW emission, is given by:
\begin{equation}
\frac{dE_{\rm gw}}{d \nu}=K \nu^3 (1+\frac{K}{\pi^2 I_{zz}} \nu^2)^{-1} \,\  \mathrm{with} \,\  \nu \in \lbrack 0-2/P_0 \rbrack,
\label{eq-Enj_pulsar}
\end{equation}
where
\begin{equation}
K = \frac{192 \pi^4 G I^3}{5 c^5 R^6} \frac{\varepsilon^2}{B^2}.
\label{eq-K_pulsar}
\end{equation}
$R$ is the radius of the star, $\varepsilon=(I_{xx}-I_{yy})/I_{zz}$ the ellipticity, $I_{ij}$ the principal moment of inertia, $B$ the projection of the magnetic dipole in the direction orthogonal to the rotation axis.
The evolution of the massive stars that give birth to pulsar being very fast, the rate can be derived directly from the star formation rate (see Eq.~(\ref{eq-rate})). Considering the interval $8-40$\,M$_\odot$ for the mass range of neutron star progenitors, and the initial mass function of Equation~(\ref{eq-imf}), \cite{reg08} found $\lambda \sim 10^{-2}$\,M$_\odot^{-1}$.

Normal radio pulsars, which are born with magnetic fields of the order of $10^{12}-10^{13}$\,G, rotational periods of the order of tens or hundreds of millisecond \citep{reg00,fau06,sor08}, are not expected to contribute significantly to the GW signal \citep{reg01b}. However the population of newborn magnetars in which super-strong crustal magnetic fields  ($B \sim 10^{14}-10^{16}$ G) may have been formed by dynamo action in a proto-neutron star with very small rotational period (of the order of $1\,$ms) \citep{dun92,tho93}, may produce a strong stochastic background in the frequency band of terrestrial detectors \citep{reg06a}. 
For these highly magnetized neutron stars,  the distortion induced by the magnetic torque becomes significant, strongly enhancing the GW emission. In the case of  a pure poloidal internal magnetic field matching to the dipolar field $B$ in the exterior, the ellipticity is given by \citep{bon96,kon00}:
\begin{equation}
\varepsilon_B=\beta \frac{R^8B^2}{4GI_{zz}^2},
\label{eq-epsB_pol}
\end{equation}
where  $\beta$ is a distortion parameter which depends on both the equation of state and
the magnetic field geometry. Using numerical simulations, \cite{bon96}  found that $\beta$ can range between $1-10$ for a
non-superconducting interior to $100-1000$ for a type I superconductor and even take values larger than $1000-10\,000$ for a type II superconductor with counter rotating electric currents.
Taking $R=10$\,km for the radius, $I_{zz}=10^{45}$\,g\,cm$^2$ for the moment of inertia, and assuming that magnetars represent $10\%$ of the population of NSs \citep{kou98}, we find that the stochastic signal is detectable with the Einstein Telescope after an observation time $T=1$\,yr and with a signal to noise ratio of 3 when $\frac{\varepsilon}{B}>1.5 \times 10^{-18}$, giving the opportunity to put very interesting constraints on both $B$ and  $\beta$.
On the other hand, It has been suggested that the spindown could become purely gravitational if the internal magnetic field could is dominated by a very strong toroidal component \citep{cut02,ste05}, of the order of $10^{16}$\,G).  In this saturation regime, the energy density increases as $\nu_{\rm o}^2$ at low frequencies and reaches a maximum of $\Omega_{\rm gw} \sim 2.3 \times 10^{-8}$ around 760\,Hz, giving a signal detectable by the Einstein Telescope with a signal-to-noise ratio of about 100 .

\subsection{Rotating Neutron Stars: Initial Instabilities}
\subsubsection{Dynamical bar modes}
The gravitational stochastic background from core collapse supernovae could be enhanced by a number of proposed post-collapse emission mechanisms. One intriguing mechanism is the bar-mode dynamical instability associated with neutron star formation. These instabilities derive their name from the `bar-like' deformation they induce, transforming a disk-like body into an elongated bar that tumbles end-over-end. The highly non-axisymmetric structure resulting from a compact astrophysical object encountering this instability makes such an object a potentially strong source of gravitational radiation and have been the subject of a number of numerical studies \citep{bro00,new00, sai00,sai01,bai07}.
Howell et al.  have calculated the background resuting signal from this emission process using simulated energy spectra data, $dE_{\rm gw}/d\nu$, from \cite{shi05}, who performed the first three dimensional hydrodynamic simulations for stellar core collapse in full general relativity. Assuming a 20\% occurrence of this instability, the authors find that the density parameter reaches a maximum of $\Omega_{\rm gw} \sim 4 \times 10^{-10}$ around 600\,Hz, and may  be detectable with the Einstein Telescope with a signal to noise ratio of 3 after one year of integration. 
The optimistic event rate considered by Howell et al. is supported by suggestions that post collapse neutrino emission by the proto-neutron stars can induce contraction through cooling. This leads to increased spins though conservation of angular momentum \citep{shi05}. The implication here is that the instability can set in tens of milliseconds post collapse, increasing the rate of occurrence.

\subsubsection{r-modes}
The stochastic background from r-modes was first investigated by \cite{owe98} and then reviewed by \cite{fer99b}. These estimates are
based on the initial model of \citep{lin98}, which does not account for
dissipation mechanisms such as the effect of the solid crust
or the magnetic field, which may significantly reduce the
gravitational instability.
The spectral energy density of a single source is given by:
\begin{equation}
\frac{dE_{\rm gw}}{d\nu }=\frac{2E_{o}}{\nu _{\sup }^2} \nu \,\ \rm{with}\,\ \nu \in \lbrack 0-\nu_{\sup}\rbrack,
\label{eq-Enj_modes}
\end{equation}
where $\nu_{\sup}$ is 4/3 of the initial rotational frequency and $E_0$ is the
rotational energy lost within the instability window.
For neutron stars with radius $R=10$ km and mass $M=1.4$ M$_\odot$ the spectrum evolves as $\Omega_{\rm gw} \sim 3 \times 10^{-12} \xi \nu_{\rm o}^3$  where $\xi$ is the fraction of NS stars born near the keplerian velocity and which enter the instability window, until it reaches a maximum at 730,Hz. The Einstein Telescope may be able to detect this signal with a signal to noise ratio larger than $> 3$ for $T=1$ yr if $\xi>0.1\%$. 
One obtains similar constraints with the secular bar mode instability at the transition between Maclaureen and Dedekind configurations \citep{lai95}.

\subsubsection{Collapse to quark matter}
It has been suggested that neutron stars could also undergo small core collapses after phase transitions,  producing large amount of gravitational waves. \cite{sig06a} calculated the background from phase transition to
quark matter in newly born NSs with millisecond periods,
based on recent numerical simulations \citep{lin06}. Assuming that 1\% of neutron stars are
born fast enough to undergo the phase transition, and that
the energy released in the process represents about  5\% of the
rotational energy ($\sim 2 \times 10^{51}$ erg), the energy density parameter
may reach a maximum of $\Omega_{\rm gw} \sim 10^{-10}$  at kHz
frequencies.

\subsection{Core Collapse Supernovas}
\subsubsection{Core collapse supernovas to neutron stars}

After they have burnt all their nuclear combustible, massive stars
may explode as type II supernovas. Their envelope is ejected while the
core collapses to form a neutron star or a black hole (BH), depending on
the initial mass of the progenitor, emitting a large amount of gravitational waves in
the process.
In a recent work, \cite{how04} calculated the stochastic background
that results from the birth of neutron stars at cosmological distances, using
relativistic numerical models of core collapse \citep{dim02}, and
updating the previous study by \cite{cow01}, based on newtonian
models \citep{zwe97}. They considered three different GW waveforms,
assumed to be representative of the three types of the catalog. Type I
waveforms are characterized by a spike resulting from the core bounce
followed by a ringdown, Type II by several distinct spikes and Type
III shows large positive and smaller negative amplitudes just before
and after bounce. In order to calculate the background spectrum, they
assumed a flat cosmology, with $\Omega_m=0.3$,
$\Omega_{\Lambda}=0.7$ and Hubble parameter $H_0=70$\,km\,s$^{-1}$\,Mpc$^{-1}$ (the 737 cosmology \citealt{rao06}), corresponding to the so-called concordant
model derived from observations of distant type Ia supernovae
\citep{per99}  and the power spectra of the cosmic microwave background
fluctuations \citep{spe03}, and considered three
different models of the SFRs, finding no sensitive difference in the
results. The NS progenitors were assumed to have masses between
$8-25$\,M$_\odot$ for a Salpeter IMF normalized between $0.1-125$\,M$_\odot$.
The background is found continuous for Type II and rather a popcorn
noise for Type I and III waveforms. The energy density parameter  reaches a
maximum of $\Omega_{\rm gw} \sim 3 \times 10^{-12}$ at around 700\,Hz for
Type I and  $\Omega_{\rm gw} \sim 10^{-13}$ at 100\,Hz and 800\,Hz
for Type II and Type III.

Besides the emission from the supernova
bounce signal in the kHz range, it is expected that the large-scale
convective overturn that develops in the delayed explosion scenario
during the epoch of shock-wave stagnation, may emit a much stronger
signal that may last for a few hundreds of ms before the actual explosion
in the 1\,Hz frequency range. \cite{buo04} estimated the background
produced by both ordinary supernovae and pop III stars using different
numerical models of the GW waveform \citep{fry04,mul04}. They showed
that the signal is Gaussian below 1\,Hz with an amplitude that may be
at the level of the background expected from inflationary models.
However, the authors stressed that these estimates remain uncertain by several
orders of magnitude, essentially due to uncertainties on the
parameters of the supernova GW emission.

\subsubsection{Core collapse supernovas to black hole}

The GW background from core collapse supernovas that result in the formation of black holes was first investigated by \cite{fer99a}, using the relativistic numerical simulations of \cite{sta85,sta86} and later by \cite{dara02a} who found similar results assuming that all the energy goes into the ringdown of the $l=m=2$ dominant quasi normal mode.
For this mode, the frequency is given by \citep{ech89}
\begin{equation}
\nu _*(m,a)  \approx \frac{\Delta(a) }{\alpha m({\rm M}_\odot)},
\end{equation}
with
\begin{equation}
\Delta(a)=\frac{c^3 }{2 \pi G} (1-0.63(1-a)^{0.3}),
\end{equation}
where $M$ is the mass of the black hole, assumed to be a fraction $\alpha$ of the mass of the progenitor $m$, and $a$ the dimensionless spin factor, ranging from 0 for a Schwarzschild BH to 1 in the extreme Kerr limit. 
The spectral energy density has the simple expression:
\begin{equation}
\frac{dE_{\rm gw}}{d\nu}=\varepsilon \alpha m c^2 \delta (\nu-\nu _*(M))
\end{equation}
where $\varepsilon$ an efficiency coefficient. Previous numerical simulations of \cite{sta85} gave an upper limit of  $\varepsilon \sim 7\times 10^{-4}$ for an axisymmetric collapse, but accounting for more realistic scenarios, in particular the pressure reduction that triggers 
the collapse. \cite{bai06} obtained  an efficiency of the order of $10^{-7}-10^{-6}$ so $2-3$ orders of magnitude smaller.
Assuming that stars in the range $30-100$\,M$_\odot$ can produce a BH, taking $\alpha=10\%$, and $a=0.6$, this simple model gives that  the energy density ranges between $0.25-5.6$\,kHz, with a maximum of $\Omega_{\rm gw} \sim \varepsilon  \times10^{-8}$ around 1650\,Hz, which means that an efficiency $>2 \times 10^{-3}$ would give a signal detectable with a signal to noise ratio of 3 after one year of observation with the Einstein Telescope. Decreasing the minimal mass or $\alpha$ would narrow the spectrum and shift the maximum toward lower frequencies, while a change in the efficiency parameter $\varepsilon$ would only affect the amplitude. Increasing the spin factor or broadening its distribution broadens the spectrum and shifts the maximum toward larger frequency. Taking $\alpha=20\%$, we find that the signal is detectable for efficiencies larger than $0.01\%$.

In a recent work, \cite{mar09} made used of the recent progress of numerical relativity, to review and extend the previous estimates of \cite{fer99a} for both population II and  population III stars. The supernova rates  were derived from the numerical simulations of \cite{tor07}, which follows the star evolution, metal enrichment and energy deposition, and the GW signal from waveform derived from relativistic numerical simulations.
The background is out of reach of the the first generation of detectors for Pop~III stellar collapse, but could be detected by the Einstein Telescope for Pop~II supernovas. 
Assuming $20-100\,\rm M_{\odot}$ for the mass range of BH progenitors,  they found that the energy density reaches a maximum  of $\Omega_{\rm gw} \sim 4-7 \times 10^{-10}$ around 500\,Hz for  the model of \cite{sek05}, giving a signal to noise ratio between $1.6-7.1$ after one year of observation.
In addition, they estimated the background from the collapse to neutron stars. Assuming $8-20\, \rm M_{\odot}$ for the mass of the progenitors, and the model of Ott (2005) which account for the g-mode excitation,  they found that the energy density reaches a maximum $\Omega_{\rm gw} \sim 10^{-9}$ around 1000\,Hz, giving a signal to noise ratio of 8.2.

Similarly, \cite{zhu10} estimated the GW signal created by all core collapse supernovae, to NS and BH, using Gaussian spectrum of the form 
\begin{equation}
\frac{dE_{\rm gw}}{d\nu}=A \exp (-(\nu-\nu_*) / 2 \sigma^2)
\end{equation}
shown to be a good approximations of the models of \cite{ott04}.
Based on simulated spectra of \cite{dim08} and \cite {sek05}, they considered different models with $\sigma \sim 500$  and $\nu_* = 200-800$ Hz. They found that the signal may be detectable for efficiencies  $\varepsilon> 10^{-5}$ and $\varepsilon> 10^{-7}$ for Einstein Telescope.
\begin{figure}
\centering
\includegraphics[angle=0,width=0.9\columnwidth]{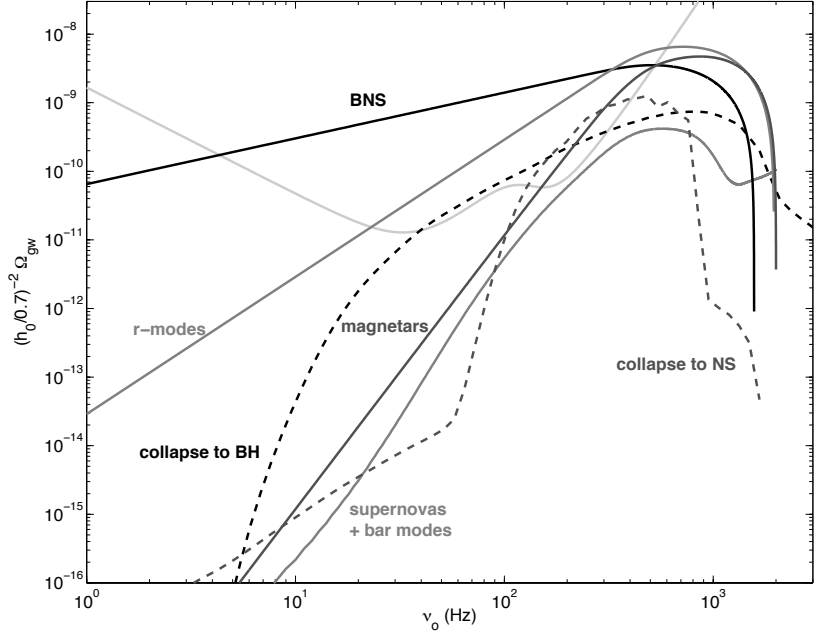}
\caption{Energy density of the most promising astrophysical background contributions for ground based detectors, discussed in the text. magnetars (threshold detectable by ET), binary neutron stars, dynamical bar modes in proto neutron stars (courtesy of E. Howell) , r-modes assuming that $1 \%$ of newborn neutron stars cross the instability window, population II core collapse to neutrons (model of \citealt{ott06}) and to black holes (model D5a of \citealt{sek05}), courtesy of S.Marrasi)
\label{fig-landscape}}
\end{figure}

\subsection{Capture by Supermassive Black Holes}
\begin{figure}[h!!!]
\centering
\includegraphics[angle=0,width=0.8\columnwidth]{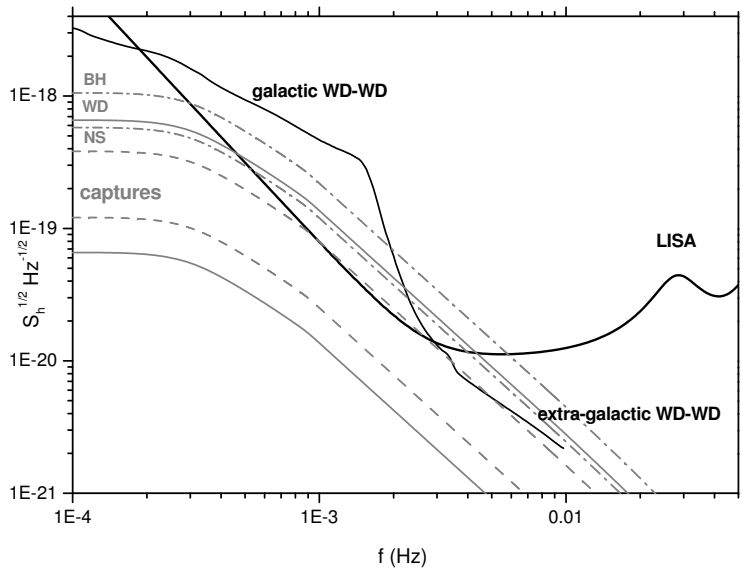}
\caption{gravitational strain in Hz$^{-1/2}$, corresponding to
  optimistic ({\it grey continuous curve}) and pessimistic ({\it grey dashed
  curve}) compact object captures \citep{bar04}, along with the LISA
  instrumental noise ({\it black}) and the WD-WD foreground ({\it black}).
\label{fig-capture}}
\end{figure}

The emission from the various populations of compact binaries, which
represent the main source of confusion noise for LISA, was studied
intensively in the past decades (see for instance \citealt{kos98,ign01,sch01,far02,coo04} for the extra-galactic contribution).
The signal is expected to be largely dominated by white
dwarf-white dwarf (WD-WD), and in particular by the galactic population
between $0.1-10$\,mHz \citep{yun00,nel01b,ben04,edl05,bel05,tim06,ben06}, the extra-galactic contribution being one order of magnitude smaller \citep{kos98,sch01}.
In a recent paper, \cite{bar04} investigated the stochastic background
created by unresolved captures by supermassive black holes (SMBHs) \citep{ama07} . The capture
rates for WDs, NSs and stellar BHs, which were extrapolated from the
rates derived by \cite{fre01} for our galaxy, represent the main source of
uncertainties, ranging between $4 \times 10^{-8}-4 \times 10^{-6}$
$M_6^{3/8}$\,yr$^{-1}$ for WD captures and between $6 \times 10^{-8}-6
\times 10^{-7}$ $M_6^{3/8}$\,yr$^{-1}$ for NS and BH captures, $M_6$
being the mass of the SMBH in units of $10^6$\,M$_\odot$.
In Figure~\ref{fig-capture}, the most optimistic and pessimistic models
are compared to the LISA instrumental noise and to the WD-WD galactic
foreground derived by \cite{ben97}.
For the most optimistic rates, the resulting  background may
contribute to the LISA confusion noise, raising the effective LISA's
overall noise level by a factor of $\sim 2$ in the range $1-10$ mHz,
where LISA is most sensitive.

\section{Conclusions and future prospects}

Gravitational Waves ground based experiments, after a decade of detector installation and commissioning, have reached or surpassed their design sensitivities, opening a new window into the Universe, The first generation LIGO interferometers have already put interesting astrophysical constraints on the ellipticity of the Crab pulsar (below the spindown limit). With advanced detectors, we expect  to see at least close compact binary coalescences, while the third generation detector Einstein Telescope and the space detector LISA should bring GW  astronomy to the next level, when it is possible to address a range of problems on a wide variety of astrophysical sources but also in fundamental physics and cosmology.

The cosmological stochastic background is often seen as the Graal of GW astronomy since it would give a snapshot of the very early stages of the Universe, up to a fraction of second after the Big Bang. The astrophysical background is also promising since it would provide information on the physical properties of compact objects  and their evolution with redshift, such as the mass of neutron stars or black holes, the ellipticity and the magnetic field of neutron stars, the angular momentum of black holes, the rate of compact binaries. We have shown in the previous sections that astrophysical models are out of reach of the first generation of detectors but with advanced detectors, and particularly the third generation Einstein Telescope, upper limits could put very interesting constraints on the equation of state and the magnetic field of magnetars, the distribution of the birth rotation period of newborn neutron stars and models of core collapse supernovas. Moreover, unless we overestimate the rate by orders of magnitude, we should be able to see the background from coalescing double neutron star binaries. 

On the other hand, the astrophysical contribution may be a noise masking the cosmological background, and also a confusion foreground where the detection of individual high redshift  standard candles needed to infer dark energy may become difficult \citep{reg09}.  In this context, modeling the astrophysical background as precisely as possible to extract informations on its strength, frequency range and statistical properties, anything that may help distinguish it from the cosmological signal or separate overlapping sources is crucial. 

Another important task in the next few years will be to adapt our actual methods to the triangle configurations of ET and the spatial interferometer LISA. It is not sure yet that we can ever get rid of the correlated noise, but we can certainly be able to reduce it by correlating interferometers formed by specific combinations of the three arms, with the extra complication for LISA that the triangle is moving.

\normalem
\begin{acknowledgements}
The author thanks Vuk Mandic for providing the landscape plot for cosmological background, Eric Howell and Stefania Marassi for their contribution to the bar mode instability and the core collapse supernovas sections.
\end{acknowledgements}


\begin{thebibliography}{142}
\providecommand{\natexlab}[1]{#1}
\providecommand{\selectlanguage}[1]{\relax}

\bibitem[{{Abbott} et~al.(2004){Abbott}, {Abbott}, {Adhikari} et~al.}]{S1}
{Abbott}, B., {Abbott}, R., {Adhikari}, R., et~al. 2004, \prd, 69, 122004

\bibitem[{{Abbott} et~al.(2005){Abbott}, {Abbott}, {Adhikari} et~al.}]{S3}
{Abbott}, B., {Abbott}, R., {Adhikari}, R., et~al. 2005, Physical Review
  Letters, 95, 221101

\bibitem[{{Abbott} et~al.(2007){Abbott}, {Abbott}, {Adhikari} et~al.}]{S4}
{Abbott}, B., {Abbott}, R., {Adhikari}, R., et~al. 2007, \apj, 659, 918

\bibitem[{{Abbott} et~al.(2009){Abbott}, {Abbott}, {Acernese} et~al.}]{S5}
{Abbott}, B.~P., {Abbott}, R., {Acernese}, F., et~al. 2009, \nat, 460, 990

\bibitem[{{Abramovici} et~al.(1992){Abramovici}, {Althouse}, {Drever}
  et~al.}]{abr92}
{Abramovici}, A., {Althouse}, W.~E., {Drever}, R.~W.~P., et~al. 1992, Science,
  256, 325

\bibitem[{{Allen}(1997)}]{all97}
{Allen}, B. 1997, in \emph{Relativistic Gravitation and Gravitational
  Radiation}, edited by {J.-A.~Marck \& J.-P.~Lasota}, 373

\bibitem[{{Allen} \& {Ottewill}(1997)}]{all97b}
{Allen}, B., \& {Ottewill}, A.~C. 1997, \prd, 56, 545

\bibitem[{{Allen} \& {Romano}(1999)}]{all99}
{Allen}, B., \& {Romano}, J.~D. 1999, \prd, 59, 102001

\bibitem[{{Amaldi} et~al.(1990){Amaldi}, {Astone}, {Bassan} et~al.}]{ast90}
{Amaldi}, E., {Astone}, P., {Bassan}, M., et~al. 1990, Europhysics Letters, 12,
  5

\bibitem[{{Amaro-Seoane} et~al.(2007)}]{ama07}
{Amaro-Seoane}, P. et~al. 2007, Classical and Quantum Gravity, 24,17
  
\bibitem[{{Ando}(2004)}]{and04}
{Ando}, S. 2004, \jcap, 6, 7

\bibitem[{{Arnaud} et~al.(1999){Arnaud}, {Cavalier}, {Davier}, \&
  {Hello}}]{arn99}
{Arnaud}, N., {Cavalier}, F., {Davier}, M., \& {Hello}, P. 1999, \prd, 59,
  082002

\bibitem[{{Astone} et~al.(1999){Astone}, {Bassan}, {Bonifazi} et~al.}]{ast99}
{Astone}, P., {Bassan}, M., {Bonifazi}, P., et~al. 1999, \aap, 351, 811

\bibitem[{{Avelino} et~al.(1998){Avelino}, {Shellard}, {Wu}, \&
  {Allen}}]{ave98}
{Avelino}, P.~P., {Shellard}, E.~P.~S., {Wu}, J.~H.~P., \& {Allen}, B. 1998,
  Physical Review Letters, 81, 2008
  
\bibitem[{{Baiotti} \& {Rezzolla}(2006){Baiotti} \& {Rezzolla}}]{bai06}
{Baiotti}, L., {de Pietri}, R., {Manca}, G.~M., \& {Rezzolla}, L. 2007, Phys. Rev. Lett, 97, 141101

\bibitem[{{Baiotti} et~al.(2007){Baiotti}, {de Pietri}, {Manca}, \&
  {Rezzolla}}]{bai07}
{Baiotti}, L., {de Pietri}, R., {Manca}, G.~M., \& {Rezzolla}, L. 2007, \prd,
  75, 044023

\bibitem[{{Ballmer}(2006)}]{bal06}
{Ballmer}, S.~W. 2006, Classical and Quantum Gravity, 23, 179

\bibitem[{{Barack} \& {Cutler}(2004)}]{bar04}
{Barack}, L., \& {Cutler}, C. 2004, \prd, 70, 122002

\bibitem[{{Battye} et~al.(1998){Battye}, {Caldwell}, \& {Shellard}}]{bat98}
{Battye}, R.~A., {Caldwell}, R.~R., \& {Shellard}, E.~P.~S. 1998, in
  \emph{Topological Defects in Cosmology}, edited by {M.~Signore \&
  F.~Melchiorri}, 11

\bibitem[{{Belczynski} et~al.(2005){Belczynski}, {Benacquista}, {Larson}, \&
  {Ruiter}}]{bel05}
{Belczynski}, K., {Benacquista}, M., {Larson}, S.~L., \& {Ruiter}, A.~J. 2005,
  arXiv:astro-ph/0510718 

\bibitem[{{Belczy{\'n}ski} \& {Kalogera}(2001)}]{bel01}
{Belczy{\'n}ski}, K., \& {Kalogera}, V. 2001, \apjl, 550, L183

\bibitem[{{Belczynski} et~al.(2006){Belczynski}, {Perna}, {Bulik}
  et~al.}]{bel06}
{Belczynski}, K., {Perna}, R., {Bulik}, T., et~al. 2006, \apj, 648, 1110

\bibitem[{{Benacquista} \& {Holley-Bockelmann}(2006)}]{ben06}
{Benacquista}, M., \& {Holley-Bockelmann}, K. 2006, \apj, 645, 589

\bibitem[{{Benacquista} et~al.(2004){Benacquista}, {DeGoes}, \&
  {Lunder}}]{ben04}
{Benacquista}, M.~J., {DeGoes}, J., \& {Lunder}, D. 2004, Classical and Quantum
  Gravity, 21, 509

\bibitem[{{Bender} \& {Hils}(1997)}]{ben97}
{Bender}, P.~L., \& {Hils}, D. 1997, Classical and Quantum Gravity, 14, 1439

\bibitem[{{Bender} \& {the LISA Study Team}(1998)}]{ben98}
{Bender}, P.~L., \& {the LISA Study Team} 1998, Laser Interferometer Space
  Antenna for the Detection of Gravitational Waves, Pre-Phase A Report, MPQ233
  (Max-Plank-Instit\"ut f\"ur Quantenoptik, Garching)

\bibitem[{{Berger} et~al.(2007){Berger}, {Fox}, {Price} et~al.}]{ber06}
{Berger}, E., {Fox}, D.~B., {Price}, P.~A., et~al. 2007, \apj, 664, 1000

\bibitem[{{Blain} et~al.(1999){Blain}, {Kneib}, {Ivison}, \& {Smail}}]{bla99}
{Blain}, A.~W., {Kneib}, J., {Ivison}, R.~J., \& {Smail}, I. 1999, \apjl, 512,
  L87

\bibitem[{{Blair} \& {Ju}(1996)}]{bla96}
{Blair}, D., \& {Ju}, L. 1996, \mnras, 283, 648

\bibitem[{{Blair} et~al.(1995){Blair}, {Ivanov}, {Tobar} et~al.}]{blai95}
{Blair}, D.~G., {Ivanov}, E.~N., {Tobar}, M.~E., et~al. 1995, Physical Review
  Letters, 74, 1908

\bibitem[{{Bonazzola} \& {Gourgoulhon}(1996)}]{bon96}
{Bonazzola}, S., \& {Gourgoulhon}, E. 1996, \aap, 312, 675

\bibitem[{{Bradaschia} et~al.(1990){Bradaschia}, {del Fabbro}, {di Virgilio}
  et~al.}]{bra90}
{Bradaschia}, C., {del Fabbro}, R., {di Virgilio}, A., et~al. 1990, Nuclear
  Instruments and Methods in Physics Research A, 289, 518

\bibitem[{{Brown}(2000)}]{bro00}
{Brown}, J.~D. 2000, \prd, 62, 084024

\bibitem[{{Buonanno}(2003)}]{buo03}
{Buonanno}, A. 2003, arXiv:gr-qc/0303085 

\bibitem[{{Buonanno} et~al.(1997){Buonanno}, {Maggiore}, \&
  {Ungarelli}}]{buo97}
{Buonanno}, A., {Maggiore}, M., \& {Ungarelli}, C. 1997, \prd, 55, 3330

\bibitem[{{Buonanno} et~al.(2005){Buonanno}, {Sigl}, {Raffelt}, {Janka}, \&
  {M{\"u}ller}}]{buo04}
{Buonanno}, A., {Sigl}, G., {Raffelt}, G.~G., {Janka}, H., \& {M{\"u}ller}, E.
  2005, \prd, 72, 084001

\bibitem[{{Caprini}(2010)}]{cap10}
{Caprini}, C. 2010, aXiv:1005.5291

\bibitem[{{Cerdonio} et~al.(1997){Cerdonio}, {Bonaldi}, {Carlesso}
  et~al.}]{cer97}
{Cerdonio}, M., {Bonaldi}, M., {Carlesso}, D., et~al. 1997, Classical and
  Quantum Gravity, 14, 1491

\bibitem[{{Cole} et~al.(2001){Cole}, {Norberg}, {Baugh} et~al.}]{col03}
{Cole}, S., {Norberg}, P., {Baugh}, C.~M., et~al. 2001, \mnras, 326, 255

\bibitem[{{Cooray}(2004)}]{coo04}
{Cooray}, A. 2004, \mnras, 354, 25

\bibitem[{{Cornish}(2001)}]{cor01}
{Cornish}, N.~J. 2001, Classical and Quantum Gravity, 18, 4277

\bibitem[{{Coward} \& {Regimbau}(2006)}]{cow06}
{Coward}, D., \& {Regimbau}, T. 2006, \nar, 50, 461

\bibitem[{{Coward} \& {Burman}(2005)}]{cow05}
{Coward}, D.~M., \& {Burman}, R.~R. 2005, \mnras, 361, 362

\bibitem[{{Coward} et~al.(2001){Coward}, {Burman}, \& {Blair}}]{cow01}
{Coward}, D.~M., {Burman}, R.~R., \& {Blair}, D.~G. 2001, \mnras, 324, 1015

\bibitem[{{Coward} et~al.(2002){Coward}, {Burman}, \& {Blair}}]{cow02a}
{Coward}, D.~M., {Burman}, R.~R., \& {Blair}, D.~G. 2002, \mnras, 329, 411

\bibitem[{{Cutler}(2002)}]{cut02}
{Cutler}, C. 2002, \prd, 66, 084025

\bibitem[{{Cyburt} et~al.(2005){Cyburt}, {Fields}, {Olive}, \&
  {Skillman}}]{cyb05}
{Cyburt}, R.~H., {Fields}, B.~D., {Olive}, K.~A., \& {Skillman}, E. 2005,
  Astroparticle Physics, 23, 313


\bibitem[{{Damour} \& {Vilenkin}(2000)}]{dam00}
{Damour}, T., \& {Vilenkin}, A. 2000, Physical Review Letters, 85, 3761

\bibitem[{{Damour} \& {Vilenkin}(2001)}]{dam01}
{Damour}, T., \& {Vilenkin}, A. 2001, \prd, 64, 064008

\bibitem[{{Damour} \& {Vilenkin}(2005)}]{dam05}
{Damour}, T., \& {Vilenkin}, A. 2005, \prd, 71, 063510

\bibitem[{{de Araujo} et~al.(2000){de Araujo}, {Miranda}, \& {Aguiar}}]{dara00}
{de Araujo}, J.~C.~N., {Miranda}, O.~D., \& {Aguiar}, O.~D. 2000, Nuclear
  Physics B Proceedings Supplements, 80, C702

\bibitem[{{de Araujo} et~al.(2002{\natexlab{a}}){de Araujo}, {Miranda}, \&
  {Aguiar}}]{dara02a}
{de Araujo}, J.~C.~N., {Miranda}, O.~D., \& {Aguiar}, O.~D. 2002{\natexlab{a}},
  Classical and Quantum Gravity, 19, 1335

\bibitem[{{de Araujo} et~al.(2002{\natexlab{b}}){de Araujo}, {Miranda}, \&
  {Aguiar}}]{dara02b}
{de Araujo}, J.~C.~N., {Miranda}, O.~D., \& {Aguiar}, O.~D. 2002{\natexlab{b}},
  \mnras, 330, 651

\bibitem[{{de Araujo} et~al.(2004){de Araujo}, {Miranda}, \& {Aguiar}}]{dara04}
{de Araujo}, J.~C.~N., {Miranda}, O.~D., \& {Aguiar}, O.~D. 2004, \mnras, 348,
  1373


\bibitem[{{de Freitas Pacheco} et~al.(2006){de Freitas Pacheco}, {Regimbau},
  {Vincent}, \& {Spallicci}}]{dfp06}
{de Freitas Pacheco}, J.~A., {Regimbau}, T., {Vincent}, S., \& {Spallicci}, A.
  2006, International Journal of Modern Physics D, 15, 235

\bibitem[{{Dimmelmeier} et~al.(2002){Dimmelmeier}, {Font}, \&
  {M{\"u}ller}}]{dim02}
{Dimmelmeier}, H., {Font}, J.~A., \& {M{\"u}ller}, E. 2002, \aap, 393, 523

\bibitem[{{Dimmelmeier} et~al.(2008){Dimmelmeier},  {Ott}, {Marek} \& {Janka}, H.-T.}]{dim08}
{Dimmelmeier}, H., {Ott}, C.~D., {Marek}, A. \& {Janka}, H.-T. 2008, \prd, 78, 064056

\bibitem[{{Drasco} \& {Flanagan}(2003)}]{dra03}
{Drasco}, S., \& {Flanagan}, {\'E}.~{\'E}. 2003, \prd, 67, {082003}

\bibitem[{{Duncan} \& {Thompson}(1992)}]{dun92}
{Duncan}, R.~C., \& {Thompson}, C. 1992, \apjl, 392, L9

\bibitem[{{Echeverria}(1989)}]{ech89}
{Echeverria}, F. 1989, \prd, 40, 3194

\bibitem[{{Edlund} et~al.(2005){Edlund}, {Tinto}, {Kr{\'o}lak}, \&
  {Nelemans}}]{edl05}
{Edlund}, J.~A., {Tinto}, M., {Kr{\'o}lak}, A., \& {Nelemans}, G. 2005, \prd,
  71, 122003


\bibitem[{{Fardal} et~al.(2007){Fardal}, {Katz}, {Weinberg}, \&
  {Dav{\'e}}}]{far07}
{Fardal}, M.~A., {Katz}, N., {Weinberg}, D.~H., \& {Dav{\'e}}, R. 2007, \mnras,
  379, 985

\bibitem[{{Farmer} \& {Phinney}(2002)}]{far02}
{Farmer}, A.~J., \& {Phinney}, E.~S. 2002, in \emph{Bulletin of the American
  Astronomical Society}, \emph{Bulletin of the American Astronomical Society},
  vol.~34, 1225

\bibitem[{{Faucher-Gigu{\`e}re} \& {Kaspi}(2006)}]{fau06}
{Faucher-Gigu{\`e}re}, C., \& {Kaspi}, V.~M. 2006, \apj, 643, 332

\bibitem[{{Ferrari} et~al.(1999{\natexlab{a}}){Ferrari}, {Matarrese}, \&
  {Schneider}}]{fer99a}
{Ferrari}, V., {Matarrese}, S., \& {Schneider}, R. 1999{\natexlab{a}}, \mnras,
  303, 258

\bibitem[{{Ferrari} et~al.(1999{\natexlab{b}}){Ferrari}, {Matarrese}, \&
  {Schneider}}]{fer99b}
{Ferrari}, V., {Matarrese}, S., \& {Schneider}, R. 1999{\natexlab{b}}, \mnras,
  303, 258

\bibitem[{{Flanagan}(1993)}]{fla93}
{Flanagan}, E.~E. 1993, \prd, 48, 2389

\bibitem[{{Freitag}(2003)}]{fre01}
{Freitag}, M. {2003}, \apjl, 583, L21

\bibitem[{{Fryer} et~al.(2004){Fryer}, {Holz}, \& {Hughes}}]{fry04}
{Fryer}, C.~L., {Holz}, D.~E., \& {Hughes}, S.~A. 2004, \apj, 609, 288

\bibitem[{{Gasperini} \& {Veneziano}(1993)}]{gas93}
{Gasperini}, M., \& {Veneziano}, G. 1993, Astroparticle Physics, 1, 317

\bibitem[{{Gasperini} \& {Veneziano}(2003)}]{gas03}
{Gasperini}, M., \& {Veneziano}, G. 2003, \physrep, 373, 1

\bibitem[{{Grishchuk}(1974)}]{gri74}
{Grishchuk}, L.~P. 1974, Soviet Journal of Experimental and Theoretical
  Physics, 40, 409

\bibitem[{{Grishchuk}(1993)}]{gri93}
{Grishchuk}, L.~P. 1993, Classical and Quantum Gravity, 10, 2449

\bibitem[{{Grishchuk} et~al.(2001){Grishchuk}, {Lipunov}, {Postnov},
  {Prokhorov}, \& {Sathyaprakash}}]{gri01}
{Grishchuk}, L.~P., {Lipunov}, V.~M., {Postnov}, K.~A., {Prokhorov}, M.~E., \&
  {Sathyaprakash}, B.~S. 2001, Physics Uspekhi, 44, 1


\bibitem[{{Hopkins} \& {Beacom}(2006)}]{hop06}
{Hopkins}, A.~M., \& {Beacom}, J.~F. 2006, {\apj}, 651, 142

\bibitem[{{Hough}(1992)}]{hou92}
{Hough}, J. 1992, in \emph{Marcel Grossmann Meeting on General Relativity},
  192--194

\bibitem[{{Hough} et~al.(1975){Hough}, {Pugh}, {Bland}, \& {Drever}}]{hou75}
{Hough}, J., {Pugh}, J.~R., {Bland}, R., \& {Drever}, R.~W.~P. 1975, \nat, 254,
  498

\bibitem[{{Howell} et~al.(2004){Howell}, {Coward}, {Burman}, {Blair}, \&
  {Gilmore}}]{how04}
{Howell}, E., {Coward}, D., {Burman}, R., {Blair}, D., \& {Gilmore}, J. 2004,
  \mnras, 351, 1237

\bibitem[{{Ignatiev} et~al.(2001){Ignatiev}, {Kuranov}, {Postnov}, \&
  {Prokhorov}}]{ign01}
{Ignatiev}, V.~B., {Kuranov}, A.~G., {Postnov}, K.~A., \& {Prokhorov}, M.~E.
  2001, \mnras, 327, 531

\bibitem[{{Jenet} et~al.(2005){Jenet}, {Hobbs}, {Lee}, \& {Manchester}}]{jen05}
{Jenet}, F.~A., {Hobbs}, G.~B., {Lee}, K.~J., \& {Manchester}, R.~N. 2005,
  \apjl, 625, L123

\bibitem[{{Kalogera} et~al.(2004){Kalogera}, {Kim}, {Lorimer} et~al.}]{kal04}
{Kalogera}, V., {Kim}, C., {Lorimer}, D.~R., et~al. 2004, \apjl, 614, L137

\bibitem[{{Kalogera} et~al.(2001){Kalogera}, {Narayan}, {Spergel}, \&
  {Taylor}}]{kal01}
{Kalogera}, V., {Narayan}, R., {Spergel}, D.~N., \& {Taylor}, J.~H. 2001, \apj,
  556, 340

\bibitem[{{Kaspi} et~al.(1994){Kaspi}, {Taylor}, \& {Ryba}}]{kas94}
{Kaspi}, V.~M., {Taylor}, J.~H., \& {Ryba}, M.~F. 1994, \apj, 428, 713

\bibitem[{{Konno} et~al.(2000){Konno}, {Obata}, \& {Kojima}}]{kon00}
{Konno}, K., {Obata}, T., \& {Kojima}, Y. 2000, \aap, 356, 234

\bibitem[{{Kosenko} \& {Postnov} (1998)}]{kos98}
{Kosenko}, D.~I. \& {Postnov}, K.~A. 1998, A\&A, 336, 736

\bibitem[{{Kopparapu} et~al.(2008){Kopparapu}, {Hanna}, {Kalogera}
  et~al.}]{kop08}
{Kopparapu}, R.~K., {Hanna}, C., {Kalogera}, V., et~al. 2008, \apj, 675, 1459

\bibitem[{{Kouveliotou} et~al.(1998){Kouveliotou}, {Dieters}, {Strohmayer}
  et~al.}]{kou98}
{Kouveliotou}, C., {Dieters}, S., {Strohmayer}, T., et~al. 1998, \nat, 393, 235

\bibitem[{{Kuroda} et~al.(2006){Kuroda}, {Kanda}, {Ohashi} et~al.}]{kur06}
{Kuroda}, K., {Kanda}, N., {Ohashi}, M., et~al. 2006, Progress of Theoretical
  Physics Supplement, 163, 54

\bibitem[{{Lai} \& {Shapiro}(1995)}]{lai95}
{Lai}, D., \& {Shapiro}, S.~L. 1995, \apj, 442, 259

\bibitem[{{Lin} et~al.(2006){Lin}, {Cheng}, {Chu}, \& {Suen}}]{lin06}
{Lin}, L., {Cheng}, K.~S., {Chu}, M., \& {Suen}, W. 2006, \apj, 639, 382

\bibitem[{{Lindblom} et~al.(1998){Lindblom}, {Owen}, \& {Morsink}}]{lin98}
{Lindblom}, L., {Owen}, B.~J., \& {Morsink}, S.~M. 1998, Physical Review
  Letters, 80{, 4843}

\bibitem[{{Lipunov} et~al.(1995){Lipunov}, {Postnov}, {Prokhorov}, {Panchenko},
  \& {Jorgensen}}]{lip95}
{Lipunov}, V.~M., {Postnov}, K.~A., {Prokhorov}, M.~E., {Panchenko}, I.~E., \&
  {Jorgensen}, H.~E. 1995, \apj, 454, 593

\bibitem[{{Lommen} et~al.(2003){Lommen}, {Backer}, {Splaver}, \&
  {Nice}}]{lom03}
{Lommen}, A.~N., {Backer}, D.~C., {Splaver}, E.~M., \& {Nice}, D.~J. 2003, in
  \emph{Radio Pulsars}, \emph{Astronomical Society of the Pacific Conference
  Series}, vol. 302, edited by {M.~Bailes, D.~J.~Nice, \& S.~E.~Thorsett},
  81

\bibitem[{{Madau} et~al.(1998){Madau}, {Pozzetti}, \& {Dickinson}}]{mad98}
{Madau}, P., {Pozzetti}, L., \& {Dickinson}, M. 1998, \apj, 498, 106

\bibitem[{{Maggiore}(2000)}]{mag00}
{Maggiore}, M. 2000, \physrep, 331, 283

\bibitem[{{Manchester}(2006)}]{man06}
{Manchester}, R.~N. 2006, Chinese Journal of Astronomy and Astrophysics
  Supplement, 6, 139

\bibitem[{{Mandic} \& {Buonanno}(2006)}]{mand06}
{Mandic}, V., \& {Buonanno}, A. 2006, \prd, 73, 063008

\bibitem[{{Marassi} et~al.(2009){Marassi}, {Schneider}, \& {Ferrari}}]{mar09}
{Marassi}, S., {Schneider}, R., \& {Ferrari}, V. 2009, \mnras, 398, 293

\bibitem[{{Mauceli} et~al.(1996){Mauceli}, {Geng}, {Hamilton} et~al.}]{mau96}
{Mauceli}{, E., {Geng}, Z.~K., {Hamilton}, W.~O.,} et~al. 1996, {\prd, 54, 1264}

\bibitem[{{Mitra} et~al.(2008){Mitra}, {Dhurandhar}, {Souradeep}
  et~al.}]{mit08}
{Mitra}, S., {Dhurandhar}, S., {Souradeep}, T., et~al. 2008, \prd, 77, 042002

\bibitem[{{M{\"u}ller} et~al.(2004){M{\"u}ller}, {Rampp}, {Buras}, {Janka}, \&
  {Shoemaker}}]{mul04}
{M{\"u}ller}, E., {Rampp}, M., {Buras}, R., {Janka}, H., \& {Shoemaker}, D.~H.
  2004, \apj, 603, 221

\bibitem[{{Nagamine} et~al.(2006){Nagamine}, {Ostriker}, {Fukugita}, \&
  {Cen}}]{nag06}
{Nagamine}, K., {Ostriker}, J.~P., {Fukugita}, M., \& {Cen}, R. 2006, \apj,
  653, 881

\bibitem[{{Nelemans} et~al.(2001){Nelemans}, {Yungelson}, {Portegies Zwart}, \&
  {Verbunt}}]{nel01b}
{Nelemans}, G., {Yungelson}, L.~R., {Portegies Zwart}, S.~F., \& {Verbunt}, F.
  2001, \aap, 365, 491

\bibitem[{{New} et~al.(2000){New}, {Centrella}, \& {Tohline}}]{new00}
{New}, K.~C.~B., {Centrella}, J.~M., \& {Tohline}, J.~E. 2000, \prd, 62, {064019}

\bibitem[{{O'Shaughnessy} et~al.(2008){O'Shaughnessy}, {Belczynski}, \&
  {Kalogera}}]{sha08}
{O'Shaughnessy}, R., {Belczynski}, K., \& {Kalogera}, V. 2008, \apj, 675, 566

\bibitem[{{Ott} et~al.(2006){Ott}, {Burrows}, {Dessart}, \& {Livne}}]{ott06}
{Ott}, C.~D., {Burrows}, A., {Dessart}, L., \& {Livne}, E. 2006, Physical
  Review Letters, 96, 201102
  
  \bibitem[{{Ott} et~al.(2006){Ott}, {Burrows}, {Livne} \&{Walder}}]{ott04}
{Ott}, C.~D., {Burrows}, A., {Livne}, E., \& {Walder}, R. 2004, \apj, 600, 834

\bibitem[{{Owen} et~al.(1998){Owen}, {Lindblom}, {Cutler} et~al.}]{owe98}
{Owen}, B.~J., {Lindblom}, L., {Cutler}, C., et~al. 1998, \prd, 58, 084020

\bibitem[{{Pallottino}(1997)}]{pal97}
{Pallottino}, G.~V. 1997, in \emph{Gravitational Waves: Sources and Detectors},
  edited by {I.~Ciufolini \& F.~Fidecaro}, 159

\bibitem[{{Perlmutter} et~al.(1999){Perlmutter}, {Aldering}, {Goldhaber}
  et~al.}]{per99}
{Perlmutter}, S., {Aldering}, G., {Goldhaber}, G., et~al. 1999, \apj, 517, 565

\bibitem[{{Phinney}(1991)}]{phi91}
{Phinney}, E.~S. 1991, \apjl, 380, L17

\bibitem[{{Piran}(1992)}]{pir92}
{Piran}, T. 1992, \apjl, 389, L45


\bibitem[{{Postnov} \& {Yungelson}(2006)}]{pos06}
{Postnov}, K.~A., \& {Yungelson}, L.~R. 2006, Living Reviews in Relativity, 9,
  6

\bibitem[{{Pradier} et~al.(2001){Pradier}, {Arnaud}, {Bizouard} et~al.}]{pra01}
{Pradier}, T., {Arnaud}, N., {Bizouard}, M., et~al. 2001, \prd, 63, 042002

\bibitem[{{Rao} et~al.(2006){Rao}, {Turnshek}, \& {Nestor}}]{rao06}
{Rao}, S.~M., {Turnshek}, D.~A., \& {Nestor}, D.~B. 2006, \apj, 636, 610

\bibitem[{Regimbau \& Chauvineau}(2007)]{reg07b}


\bibitem[{{Regimbau}(2007)}]{reg07}
{Regimbau}, T. 2007, \prd, 75, 043002

\bibitem[{{Regimbau} \& {de Freitas Pacheco}(2000)}]{reg00}
{Regimbau}, T., \& {de Freitas Pacheco}, J.~A. 2000, \aap, 359, 242

\bibitem[{{Regimbau} \& {de Freitas Pacheco}(2001{\natexlab{a}})}]{reg01a}
{Regimbau}, T., \& {de Freitas Pacheco}, J.~A. 2001{\natexlab{a}}, \aap, 376,
  381

\bibitem[{{Regimbau} \& {de Freitas Pacheco}(2001{\natexlab{b}})}]{reg01b}
{Regimbau}, T., \& {de Freitas Pacheco}, J.~A. 2001{\natexlab{b}}, \aap, 376,
  381

\bibitem[{{Regimbau} \& {de Freitas Pacheco}(2006{\natexlab{a}})}]{reg06a}
{Regimbau}, T., \& {de Freitas Pacheco}, J.~A. 2006{\natexlab{a}}, \aap, 447, 1

\bibitem[{{Regimbau} \& {de Freitas Pacheco}(2006{\natexlab{b}})}]{reg06b}
{Regimbau}, T., \& {de Freitas Pacheco}, J.~A. 2006{\natexlab{b}}, \apj, 642,
  455

\bibitem[{{Regimbau} \& {Hughes}(2009)}]{reg09}
{Regimbau}, T., \& {Hughes}, S.~A. 2009, \prd, 79, 062002

\bibitem[{{Regimbau} \& {Mandic}(2008)}]{reg08}
{Regimbau}, T., \& {Mandic}, V. 2008, Classical and Quantum Gravity, 25, 184018

\bibitem[{{Saijo} et~al.(2001){Saijo}, {Shibata}, {Baumgarte}, \&
  {Shapiro}}]{sai01}
{Saijo}, M., {Shibata}, M., {Baumgarte}, T.~W., \& {Shapiro}, S.~L. 2001, \apj,
  548, 919

\bibitem[{{Salpeter}(1955)}]{sal95}
{Salpeter}, E.~E. {1955}, \apj, 121, 161

\bibitem[{{Schneider} et~al.(2001){Schneider}, {Ferrari}, {Matarrese}, \&
  {Portegies Zwart}}]{sch01}
{Schneider}, R., {Ferrari}, V., {Matarrese}, S., \& {Portegies Zwart}, S.~F.
  2001, \mnras, 324, 797

\bibitem[{{Schnittman} et~al.(2006){Schnittman}, {Sigl}, \&
  {Buonanno}}]{schn06}
{Schnittman}, J., {Sigl}, G., \& {Buonanno}, A. 2006, in \emph{Laser
  Interferometer Space Antenna: 6th International LISA Symposium},
  \emph{American Institute of Physics Conference Series}, vol. 873, edited by
  {S.~M.~Merkovitz \& J.~C.~Livas}, 437--443

\bibitem[{{Sekiguchi} \& {Shibata}(2005)}]{sek05}
{Sekiguchi}, Y., \& {Shibata}, M. 2005, \prd, 71, 084013

\bibitem[{{Shibata} et~al.(2000){Shibata}, {Baumgarte}, \& {Shapiro}}]{sai00}
{Shibata}, M., {Baumgarte}, T.~W., \& {Shapiro}, S.~L. 2000, \apj, 542, 453

\bibitem[{{Shibata} \& {Sekiguchi}(2005)}]{shi05}
{Shibata}, M., \& {Sekiguchi}, Y. 2005, \prd, 71, 024014

\bibitem[{{Siemens} et~al.(2007){Siemens}, {Mandic}, \& {Creighton}}]{sie06}
{Siemens}, X., {Mandic}, V., \& {Creighton}, J. 2007, Physical Review Letters,
  98, 111101

\bibitem[{{Sigl}(2006)}]{sig06a}
{Sigl}, G. 2006, \jcap, 4, 2

\bibitem[{{Smith} et~al.(2006){Smith}, {Pierpaoli}, \& {Kamionkowski}}]{smi06}
{Smith}, T.~L., {Pierpaoli}, E., \& {Kamionkowski}, M. 2006, Physical Review
  Letters, 97, 021301

\bibitem[{{Soria} et~al.(2008){Soria}, {Perna}, {Pooley}, \& {Stella}}]{sor08}
{Soria}, R., {Perna}, R., {Pooley}, D., \& {Stella}, L. 2008, arXiv:0811.3605 

\bibitem[{{Spergel} et~al.(2003){Spergel}, {Verde}, {Peiris} et~al.}]{spe03}
{Spergel}, D.~N., {Verde}, L., {Peiris}, H.~V., et~al. 2003, \apjs, 148, 175

\bibitem[{{Stark} \& {Piran}(1985)}]{sta85}
{Stark}, R.~F., \& {Piran}, T. 1985, Physical Review Letters, 55, 891

\bibitem[{{Stark} \& {Piran}(1986)}]{sta86}
{Stark}, R.~F., \& {Piran}, T. 1986, in \emph{Fourth Marcel Grossmann Meeting
  on General Relativity}, edited by {R.~Ruffini}, 327--364

\bibitem[{{Starobinski{\v i}}(1979)}]{sta79}
{Starobinski{\v i}}, A.~A. 1979, Soviet Journal of Experimental and Theoretical
  Physics Letters, 30, 682

\bibitem[{{Steidel} et~al.(1999){Steidel}, {Adelberger}, {Giavalisco},
  {Dickinson}, \& {Pettini}}]{ste99}
{Steidel}, C.~C., {Adelberger}, K.~L., {Giavalisco}, M., {Dickinson}, M., \&
  {Pettini}, M. 1999, \apj, 519, 1

\bibitem[{{Stella} et~al.(2005){Stella}, {Dall'Osso}, {Israel}, \&
  {Vecchio}}]{ste05}
{Stella}, L., {Dall'Osso}, S., {Israel}, G.~L., \& {Vecchio}, A. 2005, \apjl,
  634, L165

\bibitem[{{Thompson} \& {Duncan}(1993)}]{tho93}
{Thompson}, C., \& {Duncan}, R.~C. 1993, \apj, {408}, 194

\bibitem[{{Timpano} et~al.(2006){Timpano}, {Rubbo}, \& {Cornish}}]{tim06}
{Timpano}, S.~E., {Rubbo}, L.~J., \& {Cornish}, N.~J. 2006, \prd, 73, 122001

\bibitem[{{Tong} \& {Zhang}(2009)}]{ton09}
{Tong}, M.~L., \& {Zhang}, Y. 2009, \prd, 80, 084022

\bibitem[{{Tornatore} et~al.(2007){Tornatore}, {Borgani}, {Dolag}, \&
  {Matteucci}}]{tor07}
{Tornatore}, L., {Borgani}, S., {Dolag}, K., \& {Matteucci}, F. 2007, \mnras,
  382, 1050

\bibitem[{{Tutukov} \& {Yungelson}(1994)}]{tut94}
{Tutukov}, A.~V., \& {Yungelson}, L.~R. 1994, \mnras, 268, 871

\bibitem[{{Vilenkin} \& {Shellard}(2000)}]{vil00}
{Vilenkin}, A., \& {Shellard}, E.~P.~S. 2000, \emph{{Cosmic Strings and Other
  Topological Defects}}

\bibitem[{{Wilkins} et~al.(2008){Wilkins}, {Trentham}, \& {Hopkins}}]{wil08}
{Wilkins}, S.~M., {Trentham}, N., \& {Hopkins}, A.~M. 2008, \mnras, 385, 687

\bibitem[{{Yungelson} et~al.(2001){Yungelson}, {Nelemans}, {Portegies Zwart},
  \& {Verbunt}}]{yun00}
{Yungelson}, L.~R., {Nelemans}, G., {Portegies Zwart}, S.~F., \& {Verbunt}, F.
  {2001, in \emph{The Influence of Binaries on Stellar Population Studies},
  \emph{Astrophysics and Space Science Library}, vol. 264, edited by
  {D.~Vanbeveren}, 339 (astro-ph/0011248)}

\bibitem[{{Zhu}, {Howell} \& {Blair}(2010)}]{zhu10}
{Zhu}, X.-J., {Howell}, E. \& {Blair}, D. 2010, \mnras, 409, 132

\bibitem[{{Zwerger} \& {Mueller}(1997)}]{zwe97}
{Zwerger}, T., \& {Mueller}, E. 1997, \aap, 320, 209

\end{thebibliography}

\end{document}